\documentclass[twocolumn,epjc3]{svjour3}          

\RequirePackage[T1]{fontenc}

\smartqed  
\RequirePackage[english]{babel}
\RequirePackage[utf8]{inputenc}
\RequirePackage{array}
\RequirePackage[makeroom]{cancel}
\RequirePackage{amsmath}
\RequirePackage{amstext} 
\RequirePackage{multirow}
\newcolumntype{C}{>{$}c<{$}}
\RequirePackage{graphicx}
\RequirePackage{mathptmx}      
\RequirePackage[numbers,sort&compress]{natbib}
\RequirePackage[colorlinks,citecolor=blue,urlcolor=blue,linkcolor=blue]{hyperref}
\RequirePackage{subfigure}
\RequirePackage{amssymb}
\hypersetup{pdfstartview=}

\journalname{Eur. Phys. J. C}

\begin{document}
\hfuzz=2pt
\title{Flavor changing neutral current decays $t\to c X$ ($X=\gamma,\,g,\, Z,\, H$) and $t\to c\bar \ell\ell $ ($\ell=\mu,\,\tau$) via scalar leptoquarks}



\author{
A. Bola\~nos\thanksref{e1,addr1,addr3}\and R. S\'anchez-V\'elez \thanksref{e2,addr2} \and
        G. Tavares-Velasco  \thanksref{e3,addr2}
}

\thankstext{e1}{azucena.bolanos@iberopuebla.mx}
\thankstext{e2}{ricsv05@icloud.com}
\thankstext{e3}{gtv@fcfm.buap.mx}

\institute{
Departamento de Ciencias e Ingenier\'ias, Universidad Iberoamericana, Boulevard del Ni\~no Poblano 2901, Reserva Territorial Atlixc\'ayotl,\\ CP 72820, San Andr\'es Cholula, Puebla, M\'exico\label{addr1} \and Tecnologico de Monterrey, Department of Science, Campus Puebla, Av. Atlixc\'ayotl 2301, CP 72453, Puebla, Puebla, M\'exico\label{addr3} \and
Facultad de Ciencias F\'isico-Matem\'aticas,
  Benem\'erita Universidad Aut\'onoma de Puebla,
 CP 72570, Puebla, Puebla., M\'exico \label{addr2}
}

\date{Received: date / Accepted: date}

\maketitle

\begin{abstract}
The flavor changing neutral current    decays  $t\to c X$ ($X=\gamma,\,g,\, Z,\, H$) and $t\to c\bar \ell\ell $ ($\ell=\mu,\,\tau$) are studied in a   renormalizable scalar leptoquark  (LQ) model with no proton decay,  where    a scalar $SU(2)$ doublet  with hypercharge  $Y=7/6$ is added to the standard model, yielding a  non-chiral   LQ $\Omega_{5/3}$.   Analytical results for the one-loop (tree-level) contributions of a scalar LQ  to the $f_i\to f_j X$ ($f_i\to f_j \bar f_m f_l$) decays, with $f_a=q_a, \ell_a$,  are presented. We consider the scenario  where $\Omega_{5/3}$  couples to the fermions of the second and third families, with its right- and left-handed couplings obeying $\lambda_R^{\ell u_i}/\lambda_L^{\ell u_i}=O(\epsilon)$, where $\epsilon$ para\-me\-trizes the relative size between these couplings. The allowed  parameter space   is then found via the current constraints on the muon $(g-2)$,  the  $\tau\to \mu\gamma$ decay,  the LHC Higgs boson data, and the direct LQ sear\-ches at the LHC.  For  $m_{\Omega_{5/3}}=1$ TeV and $\epsilon=10^{-3}$,  we find that the $t\to c X$  branching ratios are of similar size and can be as large as $10^{-8}$ in a tiny area of the parameter space, whereas  ${\rm Br}(t\to c\bar \tau\tau)$ [${\rm Br}(t\to c\bar \mu\mu)$]  can be up to $10^{-6}$ ($10^{-7}$).
 \end{abstract}

\section{Introduction}
The conjecture that lepton number is the fourth color quantum number was  put forward long ago  in the context of an $SU(4)_R\times SU(4)_L\times SU(4')$   theory \cite{Pati:1973uk,Pati:1974yy}, which requires the presence of new gauge and scalar bosons carrying both lepton and baryon number. Such particles, dubbed  leptoquarks (LQs) since transform  leptons into quarks and vice versa, appear naturally in  grand unified theories \cite{Georgi:1974sy,Fritzsch:1974nn,Ramond:1976jg,Senjanovic:1982ex,Frampton:1989fu}, but they are also  predicted in other well motivated theories, such as technicolor \cite{Ellis:1980hz,Farhi:1980xs,Hill:2002ap}, models with composite fermions \cite{Schrempp:1984nj,Buchmuller:1985nn,Gripaios:2009dq}, superstring-inspired ${\rm E}_6$ models \cite{Witten:1985xc,Hewett:1988xc}, models with extended scalar sectors \cite{Davies:1990sc,Arnold:2013cva}, etc. However,  LQs with lepton and baryon number violating interactions can give rise to dangerous effects such as large lepton flavor violating interactions (LFV)  or tree-level-induced proton decay. The latter implies that, unless an extra symmetry is invoked to forbid diquark couplings,  LQ masses  must be as heavy as the Planck scale, thereby rendering unobservable effects on low energy processes. Therefore, only those theories with renormalizable lepton and baryon  number conserving LQ  interactions are phenomenologically appealing.  LQ phenomenology and  low-energy constraints on the parameter space of the most representative LQ models have been  widely discussed (see  \cite{Davidson:1993qk} for instance).  For a more up-to-date review  on LQ physics  we refer the reader to Ref. \cite{Dorsner:2016wpm}. It turns out that vector LQ masses and couplings are tightly constrained by experimental data, therefore the study  of scalar LQs has been favored in the literature. In this work we are  interested in a simple renormalizable LQ  model with no proton decay where the presence of relatively light scalar LQs can still be compatible with low energy constraints from experimental data \cite{Dorsner:2009cu,Arnold:2013cva,Cheung:2015yga}. In such a model,  scalar LQs are introduced in the standard model (SM) via a doublet of $SU(2)$.

In the SM  flavor changing neutral currents (FCNCs)   can arise  up to the one-loop level or  higher orders of perturbation theory, but are additionally suppressed by the so-called Glashow--Iliopoulos--Maiani (GIM) mechanism. On the oth\-er hand, the SM forbids LFV effects at any order, though experimental evidences hint that neutrinos are massive, there\-by implying that    LFV effects    should be present in nature indeed. No evidences of large FCNCs  transitions  have yet been experimentally observed,  so the search for this class of effects is a must in the physics program of any particle collider. While FCNC transition between fermions of the second and first family are considerably constrained by experimental data, transitions involving the fermions of the third and second generation have no such strong restrictions. In this regard, FCNC top quark  transitions stand out among the most widely studied processes at the CERN LHC. This stems from the fact that in spite of having negligible rates in the SM, they can have a considerable  enhancement in beyond the SM theories and   could be at the reach of detection.

The scalar particle discovered at the LHC in 2012 seems to be consistent with the SM Higgs boson, but several  of its properties remain to be tested more accurately, such as its couplings to light fermions. The LHC also offers great potential to  search for some  exotic Higgs decay channels that  are highly suppressed or forbidden in the SM. Along these lines, there has been considerably interest in the study of the LFV Higgs boson decay $H\to \bar{\ell}_i\ell_j$, which was first studied in \cite{DiazCruz:1999xe,Han:2000jz} and has been the focus of great attention recently. Although an apparent  excess of  the $H \to \tau \mu$ branching ratio,  with a significance of $2.4\,\sigma$, was observed at the LHC Run 1  \cite{Khachatryan:2015kon}, it was not  confirmed in Run-2 data.  The LFV Higgs boson decay  $H\to \bar{\ell}_i\ell_j$ has been widely studied in several extension models. In particular, in the LQ model of Ref. \cite{Cheung:2015yga},  the $H \to \tau \mu$   rate is predicted to be at the reach of experimental detection in some regions of the allowed parameter space. On the other hand, in the quark sector, apart from the FCNC Higgs boson decay into light quarks  $H\to q q'$,  the FCNC top quark decay $t\to cH$ has also been widely studied along with other  decays such as $t\to c\gamma$, $t\to c Z$ and $t\to cg$.

At hadron colliders, the dominant top quark production mechanisms are  gluon fusion and quark annihilation. The former is the main top quark production process at the LHC (about 90 \% at $\sqrt{s}=$14 TeV), with a small percentage due to  quark annihilation. The millions of yearly top quark events at the LHC would allow experimentalist to search for its FCNCs decays such as  $t\to cX$ ($V=\gamma,\,Z,\,H$), whose rates are negligibly small in the SM \cite{Eilam:1990zc,DiazCruz:1989ub, Mele:1998ag}:
\begin{align}\label{BRFCFNCdecays}
{\rm Br}^{\rm SM}(t\to c g)&= 10^{-8},\\
{\rm Br}^{\rm SM}(t\to c\gamma)&= 10^{-10},\\
{\rm Br}^{\rm SM}(t\to cZ)&=10^{-13},\\
{\rm Br}^{\rm SM}(t\to cH)&=10^{-13}.
\end{align}
These rates however can be several orders of magnitude lar\-ger in several SM extensions, such as  two-Higgs doublet models  \cite{Eilam:1990zc,Diaz:2001vj}, supersymmetric models \cite{Couture:1994rr,Li:1993mg,Lopez:1997xv,Yang:1997dk},
left-right supersymmetric models  \cite{Frank:2005vd},
extra dimensions \cite{GonzalezSprinberg:2007zz},
models with an extra neutral gauge boson
\cite{CorderoCid:2005kp}, 331 models \cite{Cortes-Maldonado:2013rca}, etc. Therefore, any experimental evidence of these FCNCs top quark and Higgs boson decay channels may shed light on the underlying fundamental theory of particle interactions.

In this work we focus on the study of the FCNC decays $t\to cX$ ($X=\gamma,\, g,\,Z,\,H$)   in a simple renormalizable scalar LQ model in which there is no proton decay induced via tree-level LQ exchange, where these processes arise at  one-loop level at the lowest order of perturbation theory.  As a by-product, we present the exact results for the one-loop LQ scalar contribution to the $H\to \bar{f}_j f_i$ decay width, which follows easily from the $f_i\to f_j H$ decay width by crossing symmetry. For completeness, we also consider in our study the tree-level FCNC decay $t\to c\bar\ell\ell$ ($\ell=\mu,\,\tau$), which in fact can have larger branching ratios than those of the one-loop induced decays.

The rest of this presentation is as follows. In Sec. \ref{Model} we briefly discuss the framework of the LQ model we are interested in. Sec. \ref{Calculation} is devoted to present the general calculation of the FCNC decay amplitudes and decay widths. For the one-loop induced decays we express the amplitudes  in terms of both Passarino-Veltman scalar functions and Feynman parameter integrals. We present a discussion on the constraints on the LQ couplings from experimental data in Sec. \ref{Bounds}, followed by the numerical analysis of the FCNCs Higgs boson transitions in Sec. \ref{NumAnalysis}. The conclusions and outlook are presented in Sec. \ref{Conclusions}. Finally, a few lengthy formulas for the loop integrals are presented in the Appendices.

\section{A simple scalar LQ model}
\label{Model}

Rather than considering a specific  theory, a convenient strategy to study the LQ  phenomenology is via a model-indepen\-dent approach  through an effective lagrangian. One can thus focus on the low energy LQ interactions and, without loss of generality, disregard the complex framework  of the ultraviolet completion, which is not relevant for the  phenomenology below the TeV scale. The most general  dimension-four $SU(3)_c\times SU(2)_L\times U(1)_Y$-invariant  effective interactions  of  scalar and vector LQs, respecting both  lepton and baryon number was first presented in \cite{Buchmuller:1986zs} and has been analyzed recently in \cite{Dorsner:2016wpm}.  In this work we consider  a simple renormalizable LQ model in which it is not necessary to invoke an extra symmetry to forbid the  proton decay.
A single $SU(2)$ doublet  with hypercharge $7/6$  is added to the SM, giving rise to two LQs with electric charges $5/3e$ and $2/3e$.  The former one is a non-chiral LQ that couples to up quarks and charged leptons, thereby  giving rise to FCNC   top quark and Higgs boson decays at the one-loop level, but also to the $t\to c\bar\ell\ell$ decay at the tree-level. The phenomenology of this model was studied in \cite{Arnold:2013cva} and bounds on its couplings to a lepton-quark pairs from the experimental constraints on the muon anomalous magnetic dipole moment and the LFV tau decay $\tau\to\mu\gamma$ were obtained in \cite{Bolanos:2013tda}. We first start by discussing the corresponding LQ couplings to quarks and leptons and afterwards we discuss the remaining interactions.

In the model we are interested in, a scalar LQ representation $R_2$ with $SU(3)\times SU(2)\times U(1)$ quantum numbers $(3,2,7/6)$ is introduced. This LQ doublet has the following renormalizable zero-fermion-number interactions \cite{Buchmuller:1986zs}
\begin{equation}
{\cal L}_{F=0}= h^{ij}_{2L}R_2^T \bar{u}^i_Ri\tau_2 L^j_L+h^{ij}_{2R}\bar{Q}^i_Le^j_R R_2+{\rm H.c.},
\end{equation}
where ${L_L^i}$ and ${Q_L^i}$ are $SU(2)_L$ left-handed lepton and quark doublets, whereas $e^i_R$ and $q^i_R$ are singlets,
with $i$ and  $j$ being generation indices.

After rotating to the LQ mass eigenstates  $\Omega_{5/3}$ and $\Omega_{2/3}$,
where the subscript denotes the electric charge in units of $e$,  we obtain the following  interaction Lagrangian
\begin{equation}
\label{LagF=0}
{\cal L}_{F=0}=\bar{e}^i\left(\lambda_{L}^{ij}P_L+ \lambda_{R}^{ij}P_R\right)u^j \Omega_{5/3}^*+\bar{e}^i \eta_R^{ij} P_R d^j\Omega_{2/3}^*+{\rm H.c.},
\end{equation}
where $P_{L,R}$ are the chiral projection operators. We are interested in the effects of the non-chiral LQ $\Omega_{5/3}$  on the FCNC decays of the top quark and the Higgs boson.   Since there are stringent constraints on the LQ couplings to the fermions of the two first families, in our study below we will consider that  $\Omega_{5/3}$    only  couples to the  second and third generation fermions.

Apart from the LQ interaction to up quarks and charged lepton pairs, which follow easily from the above expression,  for our calculation we also need the LQ couplings to both the photon and the $Z$ gauge boson, which are extracted from the LQ kinetic terms:
\begin{equation}
\label{LQKinetic}
{\cal L}_S=\frac{1}{2}(D_\mu R_2)^\dagger D_\mu R_2,
\end{equation}
where
 the $SU(2)_L\times U(1)_Y$ covariant derivative is  given by
\begin{equation}
D_\mu R_2= \left(\partial_\mu+ig\frac{\tau^i}{2} W^i+i g'\frac{7}{6}B^\mu\right)R_2.
\end{equation}

Therefore, in the mass eigenstate basis we have
\begin{eqnarray}
{\cal L}_S&\supset&  i\frac{5e}{3}
\Omega_{5/3}\overleftrightarrow{\partial_\mu} \Omega_{5/3}^*  A^\mu- \frac{ig}{c_W}g_{Z\Omega_{5/3}\Omega_{5/3}} {\Omega_{5/3}}
\overleftrightarrow{\partial_\mu} \Omega_{5/3}^* Z^\mu \nonumber\\&+&{\rm H.c.},
\end{eqnarray}
where $g_{Z\Omega_{5/3}\Omega_{5/3}}=1/2-5/3s_W^2$.

Finally, we consider the following renormalizable effective  LQ interactions to the SM Higgs doublet $\Phi$

\begin{equation}\label{HiggsLQ}
\mathcal{L} = \left( M_{R_2}^2+\lambda_{R_2}  \Phi^\dagger \Phi\right)\left(R_2^\dagger R_2 \right),
\end{equation}
where   $M_{R_2}$ is the LQ mass.
From here we obtain the Higgs boson coupling to $\Omega_{5/3}$:
\begin{equation}
\label{ScaLagI}
{\mathcal L}\supset  \lambda_{\Omega_{5/3}} v H \Omega_{5/3}^* \Omega_{5/3}.
\end{equation}

For easy reference, we also present the SM Feynman rules for the interaction of the photon and the $Z$ gauge boson with a  fermion-antifermion pair:

\begin{align}
\label{SMFeynmanRules}
\bar{f}f A_\mu &: -ie Q_f\gamma_\mu,\\
\bar{f}f Z_\mu  &:-i\frac{g}{2c_W}\gamma_\mu (g^f_L P_L+g^f_R P_R)
\end{align}
where $g_L^f=2T^3_f-2Q_f s_W^2$ and $g_R^f=-2Q_fs_W^2$, with $T^3_f=1/2(-1/2)$ for up (down) fermions and $Q_f$ the fermion char\-ge in units of that of the positron.

\begin{figure}[hbt!]
  \centering\includegraphics[width=7cm]{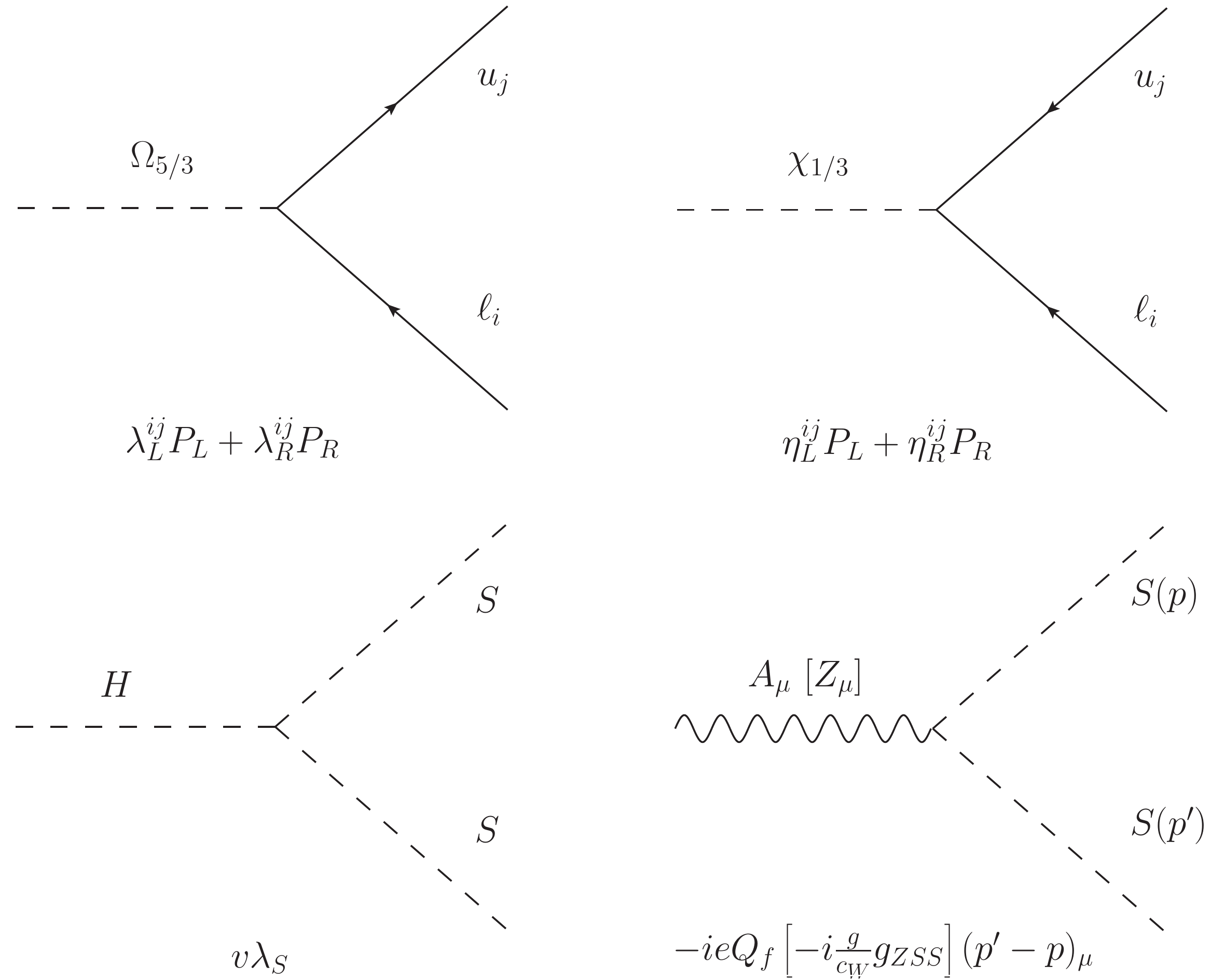}
  \caption{Feynman rules necessary for the calculation of the contribution of  the scalar LQ $\Omega_{5/3}$ to the $t\to c X$ ($X=\gamma,\,Z,\,H$) decays. For completeness we also include the Feynman rules for the LQ singlet $\chi_{1/3}$ of the model discussed in \cite{Cheung:2015yga} as our results are also valid for its contribution, thus $S$ stands for  $\Omega_{5/3}$ and $\chi_{1/3}$, with $g_{ZSS}=\frac{1}{2}-\frac{10}{3}s_W^2\;(-\frac{2}{3}s_W^2)$ for $S=\Omega_{5/3}$ ($\chi_{1/3}$).}\label{FeynmanRules}
\end{figure}

The corresponding Feynman rules follow straightforwar\-dly from the above Lagrangians and are shown in Fig.  \ref{FeynmanRules}.
Below we present the calculation of the FCNC $t\to cX$ and $t\to c\bar\ell\ell$ decays.

\section{LQ contribution to the FCNC $t\to cX $ decays}
\label{Calculation}

We now discuss the calculation of the FCNC $t\to cX$ decays, which in our scalar LQ model  proceed at the one-loop level at the lowest order of perturbation theory.    For the sake of completeness we  present the most general expressions for the $f_i\to f_j  X$  decays with $X=\gamma,\,Z,\,H$ and $f_{i,j}$ quarks  or leptons.  From our result for the $q_i\to q_j\gamma$ decay, that for the $q_i\to q_j g$ decay will follow easily as discussed below.

For the calculation of the loop integrals, we  use  both the Feynman parameter  technique and the Passarino-Veltman reduction sche\-me, which allows one to cross-check the results. For the algebra we used the Mathematica software routines along with the FeynCalc package \cite{Shtabovenko:2016sxi}.  It is worth mentioning that our results are also valid for the contribution of  the LQ singlet $\chi_{1/3}$ of the model of Ref. \cite{Cheung:2015yga}, where the LQ contribution to the $H\to \mu\tau$ was discussed. The  Feynman rules for such a LQ  are of Majorana-like type (there are two fermion-flow arrows clashing into a vertex as shown in Fig. \ref{FeynmanRules}) and  they require a special treatment. We have followed the approach of  Ref. \cite{Moore:1984eg} and   found that the results for the contribution of LQ $\Omega_{5/3}$ to the  $f_i\to f_j X$ decays  are also valid for the contributions of  LQ $\chi_{1/3}$ after replacing the respective coupling constants. A similar result was found for the contribution of single and  doubly charged scalars to the muon anomalous MDM \cite{Moore:1984eg}.
\subsection{$f_i\to f_j  V$ ($V=\gamma,\,Z$) decays}
\begin{figure}[hbt!]
  \centering\includegraphics[width=7cm]{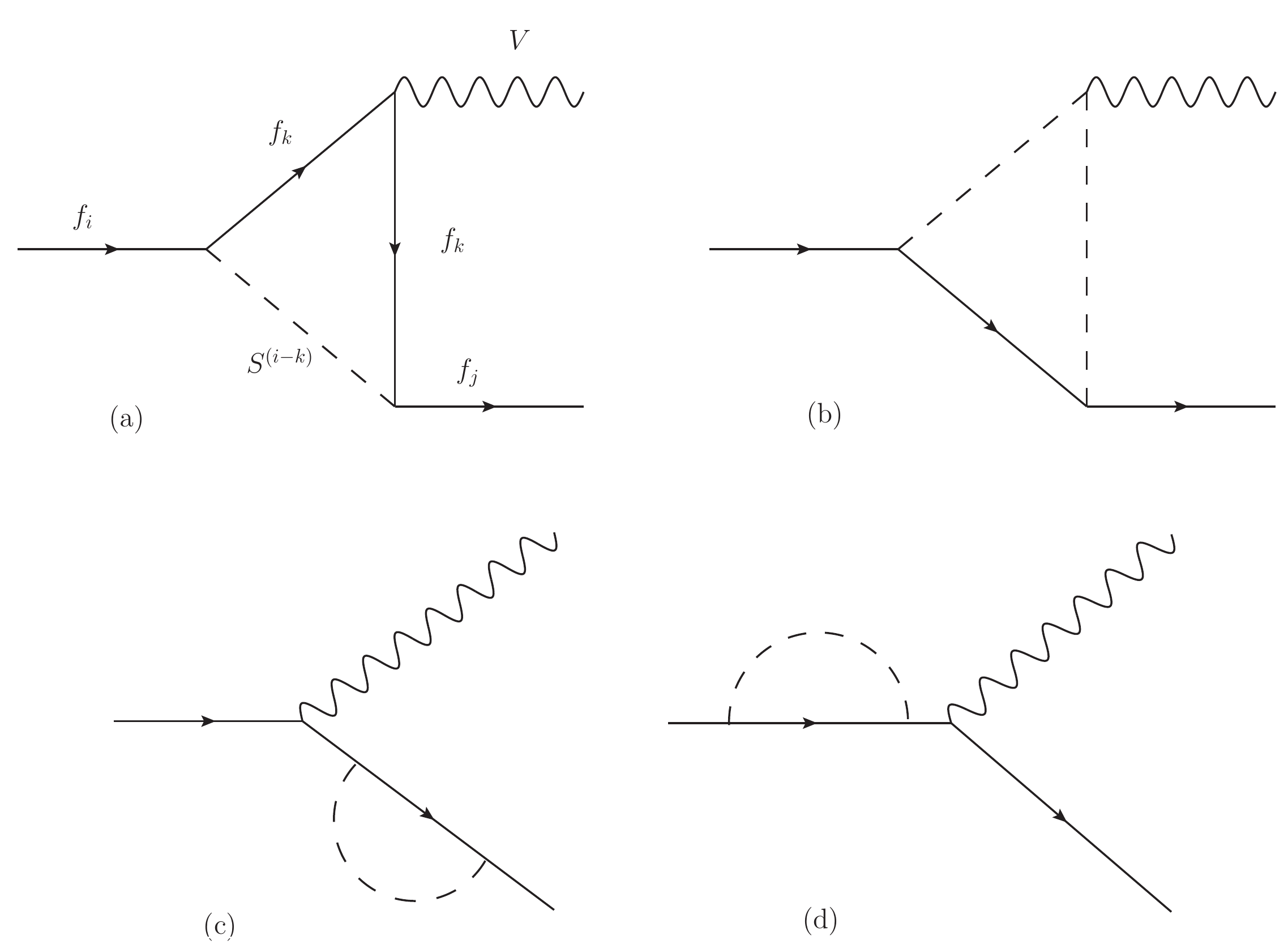}
  \caption{LQ contribution to the  decay $f_i\to f_j V$ ($V=\gamma,Z$) , where $f_i$ and $f_j$ are charged leptons (quarks) and $f_k$ is a quark (lepton). Here $Q_{i-k}$ is the LQ  charge in units of $e$.  Analogue diagrams give rise to the $f_i\to f_j H$ decay  with the $V$ gauge boson replaced by the Higgs boson. As far as the $q_i\to q_j g$ decay is concerned, there is only contribution from diagrams b) to d) as the internal fermion is a lepton.}\label{fitofjVdecay}
\end{figure}

This decay proceeds at the lowest order via the Feynman diagrams of Fig. \ref{fitofjVdecay}, where the internal fermion is a lepton (quark) provided that the external fermions are quarks (leptons).

The ultraviolet divergences cancel out when summing over all the partial amplitudes. The most general invariant amplitude can be written as
\begin{eqnarray}\label{MfitofjV}
{\cal M}(f_i\to f_j V)&=&\bar{f}_j\Big(\frac{iL^V}{m_i}\,\sigma_{\mu\nu}P_L q^\nu+\frac{iR^V}{m_i}\,\sigma_{\mu\nu}P_R q^\nu\nonumber\\&+&L^{'V}\,\gamma^\mu P_L +R^{'V}\,\gamma^\mu P_R\Big)f_i\epsilon(q)^\mu,
\end{eqnarray}
where the  monopole terms $L^{'\gamma}$ and $R^{'\gamma}$  vanish for the $f_i \to f_j \gamma$ decay due to gauge invariance: the bubble diagrams  only  give contributions to the monopole terms, which are canceled out by  those arising from the triangle diagrams.  The corresponding $L^V$, $R^V$, $L^{'Z}$, and $R^{'Z}$ form factors are presented in terms of Pa\-ssa\-rino-Veltman scalar functions in  \ref{fitofjVCoefficients}.

After averaging (summing) over  polarizations of the initial (final) fermion and  gauge boson, we use  the respective two-body decay width formula, which reduces to

\begin{align}
\label{fitofjVdecaywidth}
\Gamma(f_i\to f_j V)&=\frac{\lambda^{1/2}(m_i^2,m_V^2,m_j^2) }{32 \pi m_i^3}
\Bigg(f_{ij}\left(|L^{V}|^2+|R^{V}|^2\right)\nonumber\\&+
g_{ij}\left(|L^{'V}|^2+|R^{'V}|^2\right)\nonumber\\&+3 \left(m_j^2-m_i^2+m_V^2\right)\left(L^{'V}R^{V*}+L^{V}R^{'V*}\right)
\nonumber\\&+\frac{3 m_j}{m_i} \left(m_i^2-m_j^2+m_V^2\right)\left(L^{V}L^{'V*}+R^{V}R^{'V*}\right) \nonumber\\&-\frac{3 m_j m_V^2}{m_i}{\rm Re}\left(L^V R^{V*}\right) \nonumber\\&-12 m_i m_j {\rm Re}\left(L^{'V} R^{'V*}\right)\Bigg),
\end{align}
with
$f_{ij}=\frac{1}{m_i^2}\left(2 \left(m_i^2-m_j^2\right)^2 - \left(m_i^2+m_j^2\right)m_V^2-m_V^4\right)$ and
$g_{ij}=\frac{1}{m_V^2}\left(\left(m_i^2-m_j^2\right)^2+\left(m_i^2+m_j^2)\right)m_V^2-2 m_V^4\right)$.
The so-called triangle function is given by
\begin{equation}
\label{trianglefunc}
\lambda(x,y,z)=x^2+y^2+z^2-2(xy+xz+yz).
\end{equation}

For the $f_i\to f_j\gamma$ decay, Eq. \eqref{fitofjVdecaywidth}  reduces to
\begin{align}
\label{fitofjgamma}
\Gamma(f_i\to f_j \gamma)&=\frac{ m_i}{16 \pi   } \left(1-\left(\frac{m_j}{m_i}\right)^2\right)^3\left(|L^\gamma|^2+|R^\gamma|^2\right).
\end{align}

\subsection{$q_i\to q_j g$ decay}
\begin{figure}[hbt!]
  \centering\includegraphics[width=7cm]{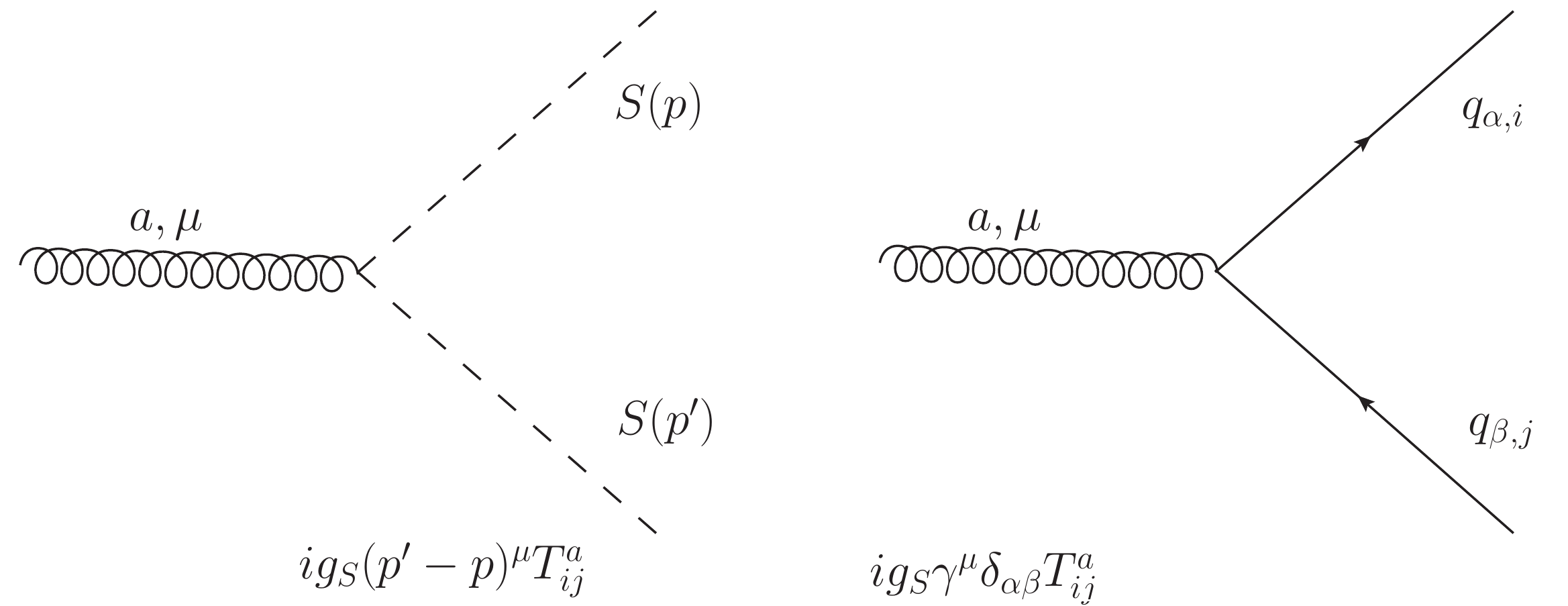}
  \caption{Feynman rules that are required for the calculation of the contribution of  a scalar LQ  to the $t\to c g$  decay. Here $T^a$ are the $SU(3)_c$ generators in the fundamental representation. }\label{FeynmanRulesgluon}
\end{figure}

This one-loop FCNC process is induced by Feynman diagrams similar to those shown in Fig. \ref{fitofjVdecay}, except that there is no contribution from Feynman diagram of type a) as  the internal fermion is a lepton.  The Feynman rules necessary for the calculation    are presented in Fig. \ref{FeynmanRulesgluon}.

The $q_i\to q_j g$ amplitude can be written as
 \begin{align}\label{Mqitoqjg}
{\cal M}(q_i\to q_j g)&=\bar{q}_jT^a\Big(\frac{iL^g}{m_i}\,\sigma_{\mu\nu}P_L q^\nu+\frac{iR^g}{m_i}\,\sigma_{\mu\nu}P_R q^\nu\Big)q_i\nonumber\\&\times\epsilon(q)_\mu,
\end{align}
 where the $L^g$ and $R^g$ coefficients  can be obtained from Eqs. \eqref{LgammaPV} and \eqref{LgammaFP} of  \ref{fitofjVCoefficients} once  the replacements   $Q_k\to 0$, $Q_S\to 1$, and $N_c e\to  g_s$ are done. After averaging (summing) over initial (final) polarizations and colors, we obtain the average square amplitude and thereby the corresponding decay width, which has the same form of Eq. \eqref{fitofjgamma}, though we must multiply the right-hand side by the color factor $C_F=4/3$.

\subsection{$f_i \to f_j H$ decay }

We now present the invariant amplitude for the  LQ contribution to the $f_i \to f_j H$ decay, which is induced at the one-loop level by   Feynman diagrams analogue to those shown in Fig. \ref{fitofjVdecay}, but with the gauge boson $V$ replaced by the Higgs boson $H$. We have found that while the amplitude of Feynman diagram (d) is ultraviolet finite, that of Feyman diagram (a) has ultraviolet divergences, but they  are canceled out by those arising from the bubble diagrams (b) and (c). After some algebra, the   invariant amplitude can be cast in the form

\begin{equation}
\label{MHtofifij}
{\cal M}(f_i \to f_j H)=\bar{f}_j\left( F_L P_L+F_R P_R\right)f_i,
\end{equation}
where the $F_L$ and $F_R$ form factors  are presented in  \ref{fitofjHCoefficients} in terms of Passarino-Veltman scalar functions and Feynman parameter integrals.

After summing (averaging) over the polarizations of the final (initial) fermion, we plug  the average squared amplitude  into the two-body decay width formula to obtain
 \begin{align}
 \label{fitofjHdecaywidth}
\Gamma(f_i \to f_j H)&= \frac{\lambda^{1/2}(m_i^2,m_H^2,m_j^2)}{8 m_i^3 \pi} \Big(\left(|F_L|^2+|F_R|^2\right) p_i \cdot p_j \nonumber \\&+2 m_i m_j{\rm Re}\left(F_L F_R^*\right)\Big),
 \end{align}
with   $p_i \cdot p_j=(m_i^2+m_j^2-m_H^2)/2$.

\subsection{$H\to \bar{f}_j f_i$ decay}
As a by-product we present the  $H\to \bar{f}_j f_i$ decay width, which follows straightforwardly from the above results by crossing symmetry. Although the  scalar LQ contribution to the LFV decay $H\to \tau \mu$  has been already presented in the zero lepton mass approximation \cite{Cheung:2015yga,Kim:2018oih,Cai:2017wry,Bauer:2015knc},  we now present the exact one-loop calculation for the $H\to \bar{f}_j f_i$ decay width. It reads

\begin{align}
\label{Htofifjdecaywidth}
\Gamma(H\to f_i f_j)&=
\frac{\lambda^{1/2}(m_H^2,m_i^2,m_j^2)}{16 \pi m_H^3}
   \Big(\left(|F_L|^2+|F_R|^2\right) p_i \cdot p_j\nonumber\\&-2 m_i
   m_j {\rm Re} (F_L F_R^*)\Big),
\end{align}
where $\Gamma(H\to f_i f_j)=\Gamma(H\to \bar{f}_i f_j)+\Gamma(H\to \bar{f}_j f_i)$. Also $p_j\cdot p_j=(m_H^2-(m_i^2+m_j^2))/2$, and the $F_L$ and $F_R$ form factors are the same as those presented in \ref{fitofjHCoefficients} for the $f_i\to f_j H$ decay as discussed in \ref{Htofifjdecay}.

\section{Three-body  tree-level decay $f_i\to f_j{\bar f}_m f_l $}

\begin{figure}[ht!]
\centering
\includegraphics[width=8cm]{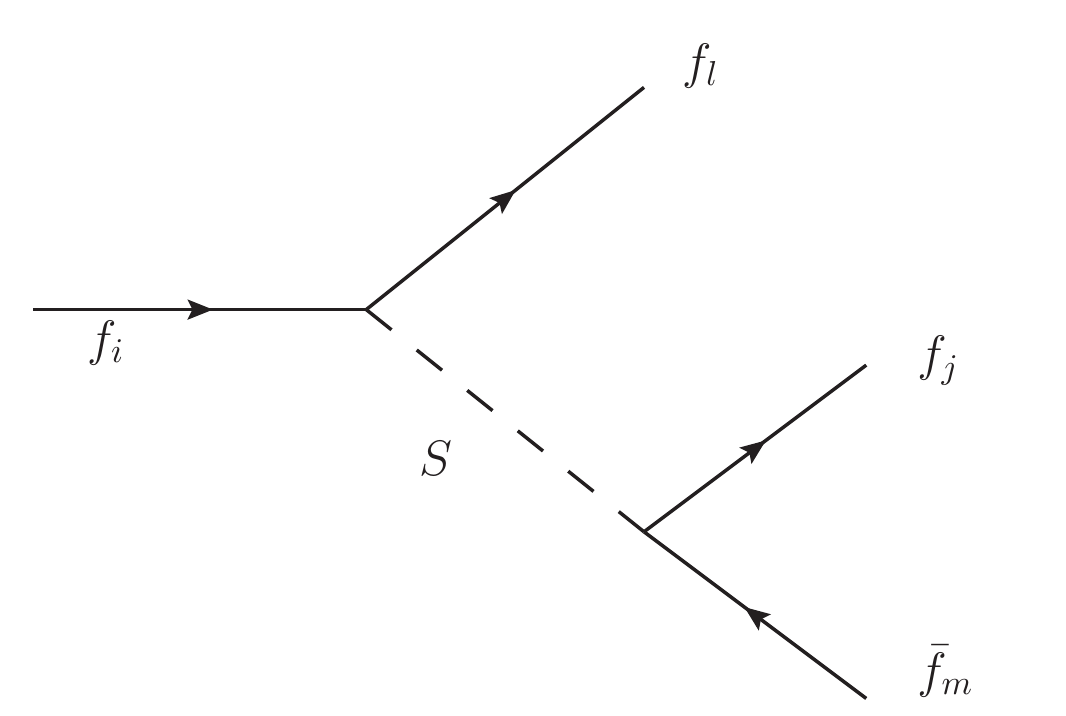}
\caption{Feynman diagram for the tree-level FCNC decay $f_i\to f_j  {\bar f}_mf_l$ induced by a scalar LQ. Here $f_l$ and $f_m$ are leptons (quarks) if $f_i$ and $f_j$ are quarks (leptons).}
\label{Fig1}
\end{figure}

Finally we discuss the calculation of the three-body decay $t\to c\bar\ell\ell$. Following our calculation approach, we consider the general decay  $f_i\to f_j  {\bar f}_mf_l$, where $f_l$ and $f_m$ are leptons (quarks) if $f_i$ and $f_j$ are quarks (leptons). This process is induced by a scalar LQ at the tree-level  via the Feynman diagram of Fig. \ref{Fig1}. We denote the  four-momentum of fermion $f_a$ ($a=i, j,l,m$) by $p_a$.
The corresponding decay width can be written as

\begin{equation}
\label{fitofjfmfldecay}
\Gamma(f_i\to f_j \bar{f}_m f_l)=\frac{m_i}{256 \pi^3}\int_{x_{ji}}^{x_{jf}}\int_{x_{li}}^{x_{lf}}|\mathcal{\overline{M}}|^2dx_j dx_l.
\end{equation}
In the center-of-mass frame of the decaying fermion,  the scaled variables  $x_{a}$ ($a=j,l,m$) are given as $x_{a}=2 E_{a}/m_{a}$.  From energy conservation, these variables obey $x_j+x_l+x_m=2$.
The kinematic limits in Eq. \eqref{fitofjfmfldecay} are in turn
\begin{eqnarray}
x_{jf}&=&2\sqrt{\mu_j},\\
x_{ji}&=&1+\mu_j-\mu_l-\mu_m-2\sqrt{\mu_l\mu_m},\\
x_{li,lf}&=&\frac{1}{2(1-x_j+\mu_j)}\Bigl[(2-x_j)(1+\mu_j+\mu_l-\mu_m-x_j)\nonumber\\
&\mp& \sqrt{x_j^2-4\mu_j}\lambda^{1/2}(1+\mu_j-x_i,\mu_l,\mu_m)\Bigr],
\end{eqnarray}
where $\mu_a=m_a^2/m_i^2$ $(a=j,l,m)$.

The square average amplitude can be expressed as

\begin{eqnarray}
|\mathcal{\overline{M}}|^2&=&\frac{2 (p_j\cdot p_m (|\lambda
   ^{jm}_L|^2+|\lambda
   ^{jm}_R|^2)-2 m_m m_j \lambda
   ^{jm}_L \lambda ^{jm}_R)}{(m_m^2+m_j^2-m_S^2-2 p_j\cdot p_m)^2}\nonumber\\&\times& \Bigl(  (m_l^2+p_l\cdot p_m+p_j\cdot p_l)(|\lambda ^{il}_L|^2+|\lambda
   ^{il}_R|^2)\nonumber\\&+&2
   m_i m_l \lambda ^{il}_L \lambda
   ^{il}_R
  \Bigr),
\end{eqnarray}
where the scalar products can be written  as

\begin{eqnarray}
p_l\cdot p_m&=&\frac{m_i^2}{2}(1 + \mu_j - \mu_l - \mu_m - x_j),\\
p_j\cdot p_m&=&\frac{m_i^2}{2}(1-\mu_j+\mu_l-\mu_m-x_l),\\
p_j\cdot p_l&=&\frac{m_i^2}{2}(1-\mu_j+\mu_m-\mu_l-x_m).
\end{eqnarray}
The integration of Eq. \eqref{fitofjfmfldecay} can be performed numerically.

\section{Constraints on the parameter space of the scalar LQ models}
\label{Bounds}
We now consider the LQ model introduced above and pre\-sent an analysis of the constraints on  the LQ  couplings to SM fermions and the Higgs boson. While the LQ couplings to fermions can be obtained from the muon anomalous MDM and LFV tau decays, the LQ coupling to a Higgs boson pair can be extracted from the constraint on the $H\gamma\gamma$ and $Hgg$ couplings obtained by the ATLAS and CMS collaborations   \cite{Khachatryan:2016vau}.

\subsection{Constraints on scalar LQ masses}

The phenomenology of  the  scalar LQ doublet  $R_2$ has been long studied in the literature
\cite{Shanker:1982nd,Davidson:1993qk,Mizukoshi:1994zy,Arnold:2013cva, Dorsner:2013tla,Bolanos:2013tda}, and constraints on their mass and couplings have been derived from the $Z\to b\bar{b}$ decay, the muon anomalous MDM, and LFV decays. Since  low energy physics strongly constrains  the LQ couplings  to the first-generation fermions,  it is usually assumed that the only non-negligible couplings are those  to the fermions of the second and third generations. The most stringent current constraint on the mass of the scalar LQ doublet    $R_2$ masses is $m_{\Omega_{2/3}}\gtrapprox 1$ TeV, which was obtained by the ATLAS \cite{Aaboud:2019bye} and CMS \cite{Sirunyan:2018vhk} collaborations from  the  LHC data at  $\sqrt{s}=13$ TeV under the  assumption that $\Omega_{2/3}$ is a third-generation LQ that decays  mainly as $\Omega_{2/3}\to \bar{\tau}b$, though such a bound relaxes up to 800 GeV when it is assumed that  $\Omega_{2/3}$ decays into both the $\bar{\tau}b$ and $t\nu_\tau$ channels.  Also, the  LQ search via pair production \cite{Sirunyan:2018ryt} gives a very stringent upper bound of $1500$ GeV on the mass of second-generation LQs, which we do not consider here as we are interested in a LQ that couples to both second and third-generation fermions.
We will then assume the less stringent bound  $m_{\Omega_{5/3}}\ge 800$ GeV in our analysis below since   $m_{\Omega_{2/3}}$ and $m_{\Omega_{5/3}}$ are mass degenerate, {\it cf.} Eq. \eqref{HiggsLQ}.  In fact, a non-degenerate scalar LQ doublet could give dangerous contributions to the oblique parameters
\cite{Keith:1997fv},

\subsection{Constraints from the LHC data on the Higgs boson}
 LHC data indicate that the 125 GeV Higgs boson couplings are compatible with those predicted by the SM, which provides a useful approach to constrain the parameter space of SM extension models by means of the so-called Higgs boson coupling modifiers, which are defined as
 \begin{equation}
 \label{Higgsmodifiers}
 \kappa_i^2=\frac{\Gamma^{}(H\to i)}{\Gamma^{\rm SM}(H\to i)},
 \end{equation}
 where $\Gamma^{\rm SM}(H\to i)$ is the SM Higgs boson decay width and $\Gamma(H\to i)$ is the one  including new physics effects.
 Bounds on  the Higgs boson coupling modifiers were obtained by fitting the combined data of the ATLAS and CMS collaborations \cite{Khachatryan:2016vau}. Since LQs contribute at the one-loop level to the $H\to\gamma\gamma$ and $H\to gg$ decays, to constrain the LQ couplings to a Higgs boson pair $H\Omega_{5/3}\Omega_{5/3}$, we  use  $\kappa_\gamma$ and $\kappa_g$,  which  are given as \cite{Dorsner:2016wpm}

\begin{equation}
\kappa_\gamma\simeq \dfrac{\left|F_1\left(\tau_W\right) +\frac{4}{3}F_{1/2}\left(\tau_t\right)+\sum\limits_i
\dfrac{3Q_{S_i}^2\lambda_{S_i}\upsilon^2}{2m_{S_i}^2}F_0(\tau_{S_i})
\right|}{\left|F_1\left(\tau_W\right) +\frac{4}{3}F_{1/2}\left(\tau_t\right)\right|},
\end{equation}
and
\begin{equation}
\kappa_g\simeq \dfrac{\left|\frac{1}{2}F_{1/2}\left(\tau_t\right)+ \sum\limits_{i}\dfrac{\lambda_{S_i}\upsilon^2}{4m_{S_i}^2} F_{0}\left(\tau_{S_i}\right)\right|}
{\left|\frac{1}{2}F_{1/2}\left(\tau_t\right)\right|},
\end{equation}
where the sum is over the LQs $S_i$,  $\tau_a= 4 m_{H}^2/m_a^2$, and the $F_{s}(\tau_a)$ function is given by
\begin{equation}
F_{s}(\tau)=\left\{
\begin{array}{lcl}
-2\tau(1+(1-\tau)f(\tau)) &&\quad s=1/2,\\ \\
2+3\tau+3\tau(2-\tau)f(\tau)&&\quad s=1,\\ \\
\tau(1-\tau f(\tau)))&&\quad s=0,
 \end{array}\right.
\end{equation}
where
\begin{equation}
f(x)=\left\{
\begin{array}{cr}
\left[\arcsin\left(\frac{1}{\sqrt{x}}\right)\right]^2&x\ge1,\\
-\frac{1}{4}\left[\log\left(\frac{1+\sqrt{1-x}}{1-\sqrt{1-x}}\right)-i\pi\right]^2&x<1.
\end{array}
\right.
\end{equation}

\begin{figure*}[!htb]
\centering\includegraphics[width=14cm]{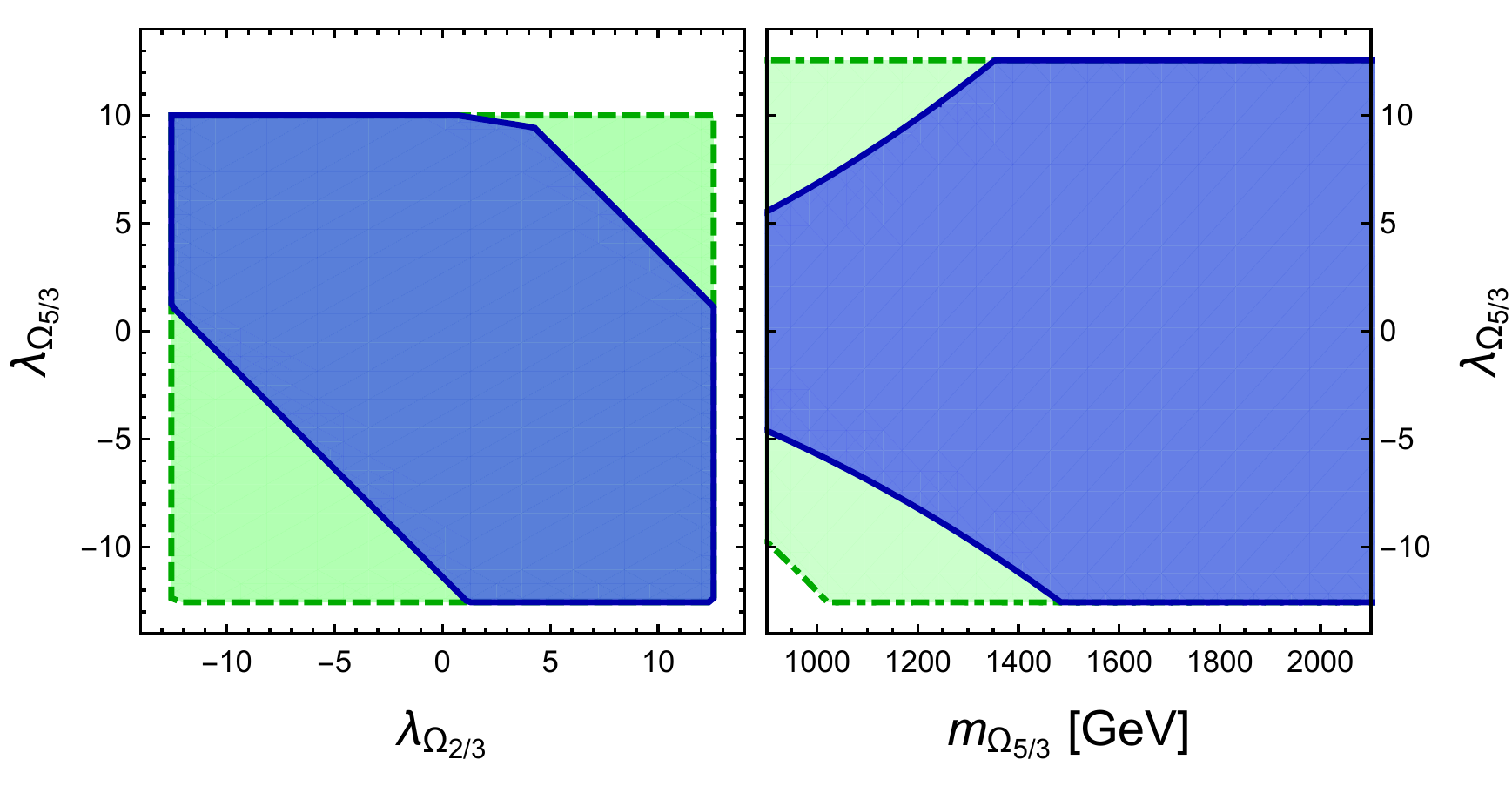}
\caption{Allowed regions with 95 \% C.L. of the parameter space of our LQ model from the experimental bounds on the Higgs boson multipliers $\kappa_\gamma$ and $\kappa_g$ for $m_{\Omega_{2/3}}=m_{\Omega_{5/3}}$.
The left plot shows the allowed region in the $\lambda_{\Omega_{5/3}}$ vs $\lambda_{\Omega_{2/3}}$ plane for three values of  $m_{\Omega_{5/3}}$: 1000 GeV (solid-line boundary), and 1500 GeV (dashed-line boundary). The right plot  shows the allowed area in the $\lambda_{\Omega_{5/3}}$ vs $m_{\Omega_{5/3}}$ plane   in the following  scenarios of $\lambda_{\Omega_{2/3}}$:  $\lambda_{\Omega_{2/3}}=\lambda_{\Omega_{5/3}}$ (solid-line boundary) and  $\lambda_{\Omega_{2/3}}=-\lambda_{\Omega_{5/3}}$ (dot-dashed-line boundary). The vertical and horizontal lines correspond to  the bounds from perturbativity $|\lambda_{\Omega_{2/3,\,5/3}}|\le 4\pi$. \label{HiggsLQCoupBoundsI}}
\end{figure*}

Although  in  our model FCNCs top quark decays receive contribution from  $\Omega_{5/3}$ only,   $\Omega_{2/3}$ also contribute to  the decays $H\to\gamma\gamma$ and $H\to gg$. As already mentioned, these LQs are mass degenerate: $m_{\Omega_{5/3}}= m_{\Omega_{2/3}}$.
We  show in the left plot of Fig.   \ref{HiggsLQCoupBoundsI}  the area allowed by the experimental constraints on  $\kappa_\gamma$ and $\kappa_g$ in the $\lambda_{\Omega_{5/3}}$ vs $\lambda_{\Omega_{2/3}}$ plane for two values of $m_{\Omega_{5/3}}$. In general, values of the order of $O(10)$ are allowed for either   $\lambda_{\Omega_{2/3}}$ or $\lambda_{\Omega_{5/3}}$, with the largest allowed values obtained for  either  large $m_{\Omega_{5/3}}$  or  $\lambda_{\Omega_{2/3}}=-\lambda_{\Omega_{5/3}}$. We also show the allowed area in the  $m_{\Omega_{5/3}}$ vs $\lambda_{\Omega_{5/3}}$ plane  in several $\lambda_{\Omega_{2/3}}$ scenarios. We observe that for a particular $m_{\Omega_{5/3}}$ value,   the strongest constraints are obtained when $\lambda_{\Omega_{2/3}}=\lambda_{\Omega_{5/3}}$, whereas the less stringent constraints are obtained when   $\lambda_{\Omega_{2/3}}=-\lambda_{\Omega_{5/3}}$.

In summary  the $\kappa_\gamma$ and $\kappa_g$ constraints are satisfied for  $\lambda_{\Omega_{5/3}}$  of the order of $O(10)$, with the largest values allowed for a heavy LQ. In our analysis below we will use however the conservative value $\lambda_{\Omega_{5/3}}\simeq 1$ as a very large value would violate the perturbativity of the LQ coupling.

\subsection{Constraints from the muon anomalous magnetic moment and the LFV decay $\tau\to\mu\gamma$}

The experimental bounds on the muon anomalous magnetic dipole moment (MDM)  $a_\mu$ and the LFV tau decays provide an useful tool to constrain  LFV effects \cite{Tanabashi:2018oca}. In particular, $a_\mu$ can be useful to constrain the  LQ couplings $\lambda^{\mu u_i}_{L,R}$ ($u_i=c,t$), whereas  the decay $\tau\to\mu\gamma$ allow us to constrain the $\lambda^{\tau u_i}_{L,R}$ ones.

\subsubsection{Muon anomalous magnetic dipole moment}

Currently there is a discrepancy between the experimental and theoretical values of the muon anomalous MDM $\Delta a_\mu=a_\mu^{\rm Theo.}-a_\mu^{\rm Exp.}=268\, (63)\,(43)\times 10^{-11}$ \cite{Tanabashi:2018oca}. We assume that this discrepancy is due to  the LQ contribution, though such a puzzle could be settled in the future once  new  experimental measurements  and more accurate evaluations of the hadronic contributions were available.

\begin{figure*}[!htb]
\centering\includegraphics[width=14cm]{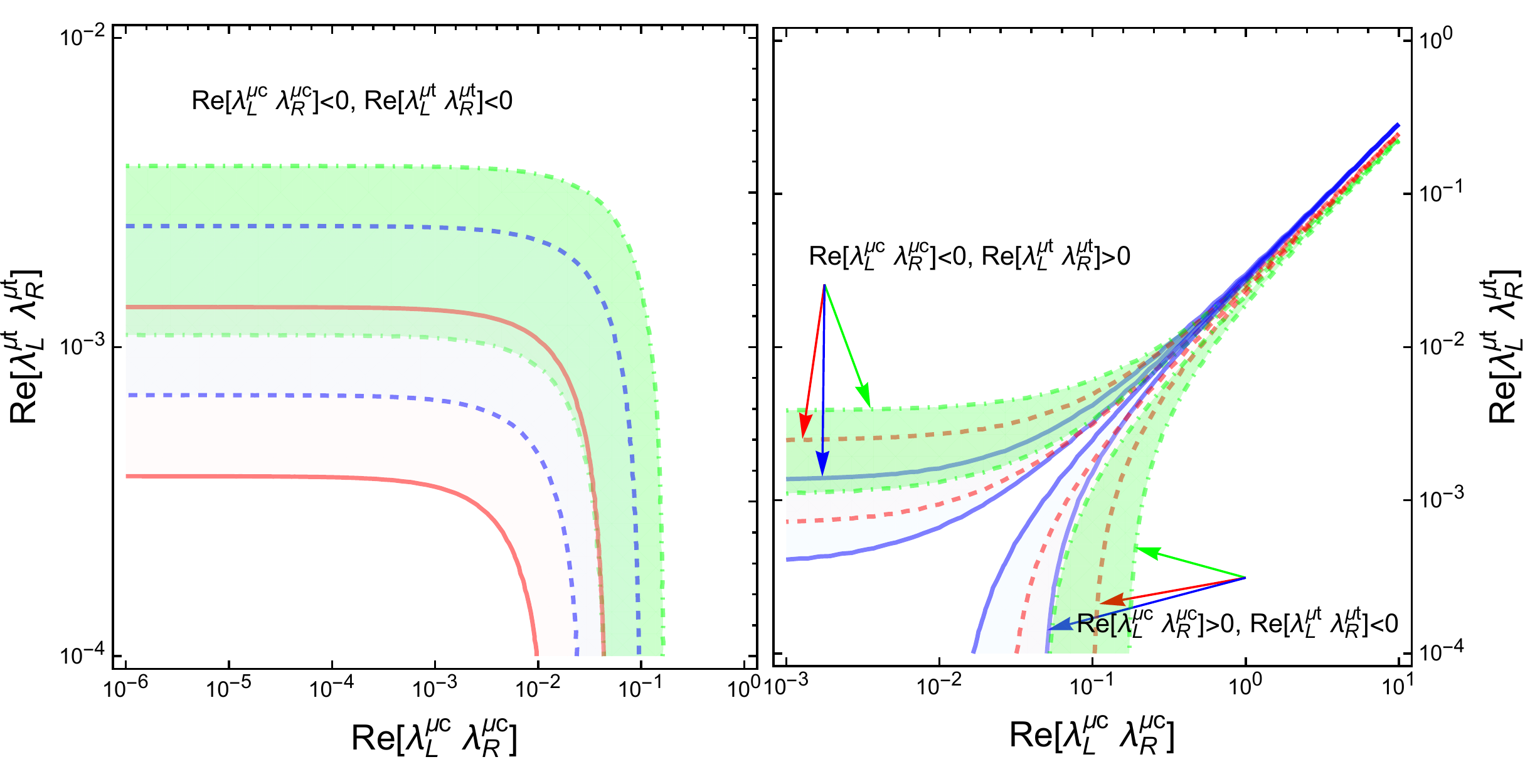}
\caption{Allowed regions with 95 \% C.L. of the parameter space of our LQ model assuming that the contributions of LQ $\Omega_{5/3}$ along with the $c$ and $t$ quarks are responsible for  the muon anomalous MDM discrepancy $\Delta a_\mu$. We show the allowed region in the   ${\rm Re}\left(\lambda_{L}^{\mu c}\lambda_{R}^{\mu c}\right)$ vs ${\rm Re}\left(\lambda_{L}^{\mu t}\lambda_{R}^{\mu t}\right)$ plane for two values of  $m_{\Omega_{5/3}}$: 1000 GeV (solid-line boundary), 1500 GeV (dashed-line boundary), and 2000 GeV (dot-dashed-line boundary). In the left plot we assume that  both ${\rm Re}\left(\lambda_{L}^{\mu c}\lambda_{R}^{\mu c}\right)<0$ and  ${\rm Re}\left(\lambda_{L}^{\mu t}\lambda_{R}^{\mu t}\right)<0$, whereas in  the right plot we consider that  they  are of opposite sign.   \label{amuBoundI}}
\end{figure*}

The   contribution of  scalar LQs $\Omega_{5/3}$  to the muon anomalous magnetic dipole
moment $a_{\mu}^{\rm LQ}$   arises at the one-loop level from the triangle diagrams of Fig. \ref{fitofjVdecay} with $f_j=f_i=\mu$ and $f_k=u_k$.  It  can be written as \cite{Bolanos:2013tda}

\begin{align}
\label{afi}
a_{\mu}^{\rm LQ}&=-\sum_ {u_k=c,\,t}\frac{3\,\sqrt{x_{\mu}}}{32\pi^2} \Bigg(\sqrt{x_\mu}\left(\left|\lambda^{\mu u_k}_L\right|^2+
\left|\lambda^{\mu u_k}_R\right|^2\right) F\left(x_\mu,x_{u_k}\right)\nonumber\\& +2\,\sqrt{x_{u_k}}\,{\rm Re}\left(\lambda^{\mu u_k}_L
{\lambda^{\mu u_k }_R}^*\right) G\left(x_\mu,x_{u_k}\right) \Bigg),
\end{align}
where $x_a=m_a^2/m^2_{\Omega_{5/3}}$.  The $F(x,y)$ and $G(x,y)$ functions are presented in \ref{AnomalousMDM} in terms of Feynman parameter integrals and Passarino-Veltman scalar functions. Since $x_\mu\ll x_{u_k}$, we have the following approximate expression
\begin{align}
\label{afiap}
a_{\mu}^{\rm LQ}&\simeq-\sum_ {u_k=c,\,t}\frac{3\,\sqrt{x_{\mu}}\sqrt{x_{u_k}}}{16\pi^2}\,{\rm Re}\left(\lambda^{\mu u_k}_L
{\lambda^{\mu u_k }_R}^*\right) G\left(x_\mu,x_{u_k}\right) ,
\end{align}
Since $\Omega_{2/3}$  is a chiral LQ, its contribution to $a_\mu$ is proportional to the muon mass and is thus subdominant. We now consider  that the $\Delta a_\mu$ discrepancy is due to  the LQ contribution $a_{\mu}^{\rm LQ}$ and show in Fig.  \ref{amuBoundI} the allowed area in the ${\rm Re}\left(\lambda_{L}^{\mu c}\lambda_{R}^{\mu c}\right)$ vs ${\rm Re}\left(\lambda_{L}^{\mu t}\lambda_{R}^{\mu t}\right)$ plane for three values of $m_{\Omega_{5/3}}$. We note that a positive contribution from LQs to $a_\mu$ is required to explain the discrepancy, therefore there are three possible scenarios:
\begin{enumerate}
\item${\rm Re}\left(\lambda_{L}^{\mu c}\lambda_{R}^{\mu c}\right)$ and ${\rm Re}\left(\lambda_{L}^{\mu t}\lambda_{R}^{\mu t}\right)<0$.
\item${\rm Re}\left(\lambda_{L}^{\mu c}\lambda_{R}^{\mu c}\right)<0$ and ${\rm Re}\left(\lambda_{L}^{\mu t}\lambda_{R}^{\mu t}\right)>0$.
\item${\rm Re}\left(\lambda_{L}^{\mu c}\lambda_{R}^{\mu c}\right)>0$ and ${\rm Re}\left(\lambda_{L}^{\mu t}\lambda_{R}^{\mu t}\right)<0$.
\end{enumerate}
In the first scenario (left plot of Fig. \ref{amuBoundI}) we observe that while ${\rm Re}\left(\lambda_{L}^{\mu c}\lambda_{R}^{\mu c}\right)$ can range between $10^{-4}$ and $10^{-3}$ for negligible ${\rm Re}\left(\lambda_{L}^{\mu c}\lambda_{R}^{\mu c}\right)$, the latter can range between $10^{-3}$ and $10^{-2}$ for negligible ${\rm Re}\left(\lambda_{L}^{\mu t}\lambda_{R}^{\mu t}\right)$, with the largest allowed values corresponding to heavy $m_{\Omega_{5/3}}$. On the other hand,  more large values of the LQ couplings are allowed when   ${\rm Re}\left(\lambda_{L}^{\mu c}\lambda_{R}^{\mu c}\right)$ and ${\rm Re}\left(\lambda_{L}^{\mu t}\lambda_{R}^{\mu t}\right)$ are of opposite sign (right plot) as there is a cancellation between the contributions of the $c$ and $t$ quarks. In particular, there is a  very narrow band where ${\rm Re}\left(\lambda_{L}^{\mu c}\lambda_{R}^{\mu c}\right)\simeq  O(10)$ and ${\rm Re}\left(\lambda_{L}^{\mu t}\lambda_{R}^{\mu t}\right)\simeq  O(1)$.

\subsubsection{Decay $\tau\to\mu\gamma$}
The LQ couplings $\lambda_{L,R}^{\tau u_i}$ and $\lambda_{L,R}^{\mu u_i}$  can  be constrained by the experimental bound on the LFV tau decay $\tau\to\mu\gamma$, which can receive the contributions of loops with  $\Omega_{5/3}$ accompanied by the up quarks. Such contributions follow straightforwardly from our result for the $f_i\to f_j\gamma$ decay width given in Eq. \eqref{fitofjVdecaywidth} after the proper replacements are made. The result  is in agreement with previous calculations of the $\ell_i\to\ell_j\gamma$ decay width \cite{Bolanos:2013tda}.

\begin{figure*}
\centering\includegraphics[width=14cm]{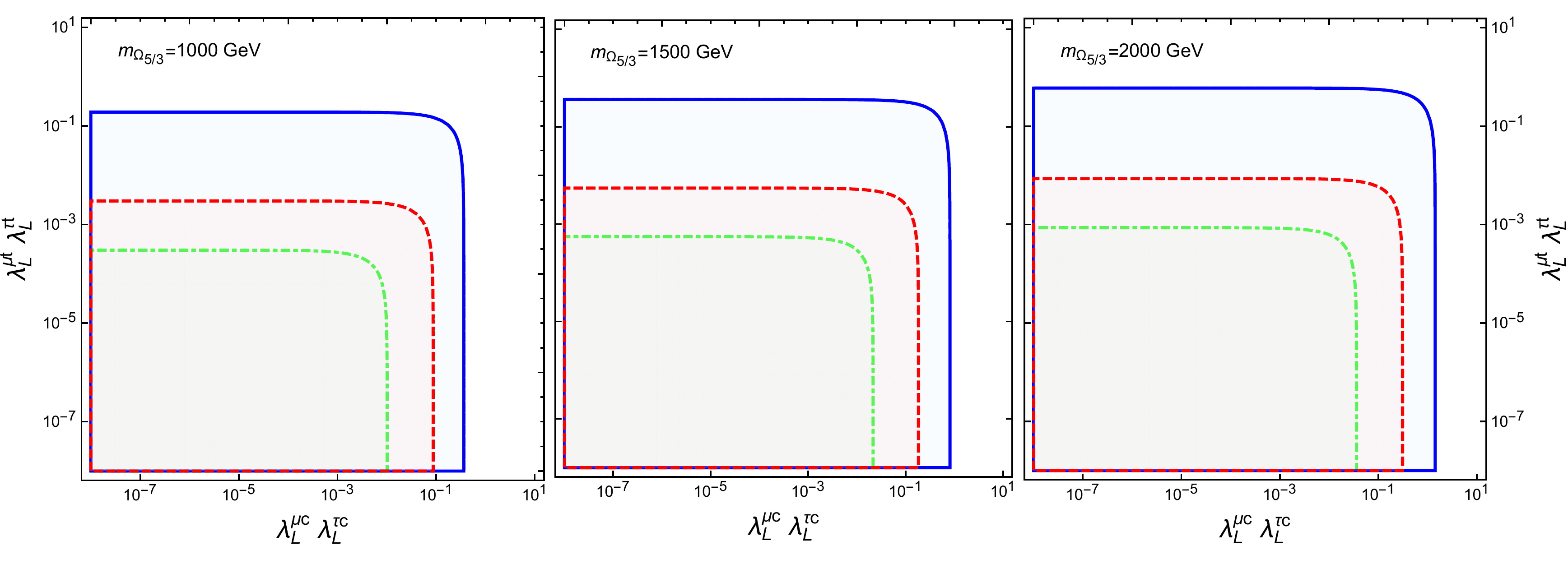}
\caption{Allowed area with 95\% C.L. in the $\lambda_L^{\mu c} \lambda_L^{\tau c}$ vs $\lambda_L^{\mu t}\lambda_L^{\tau t }$ plane from the experimental bound on the $\tau\to\mu\gamma$ decay for three values of $m_{\Omega_{5/3}}$ in the scenario with $\lambda_R^{\ell u_i}/\lambda_L^{\ell u_i}=O(\epsilon)$, for   $\epsilon=10^{-3} $ (solid-line boundary), $\epsilon=0.1$ (dashed-line boundary), and $\epsilon=1$ (dot-dashed-line boundary). \label{LFVBoundI1}}
\end{figure*}

If the LQ couples to both the $c$ and $t$ quarks, the $\tau \to \mu\gamma$ decay width acquires the form

\begin{align}
\Gamma(\tau \to \mu\gamma)&\sim \Big\|\sum_{u_i=c,t}\left(\alpha_{LL}^{\mu u_i}\lambda_L^{\mu u_i} \lambda_L^{\tau u_i} +\alpha_{RR}^{\mu u_i}\lambda_R^
{\mu u_i} \lambda_R^{\tau u_i}\right.\nonumber\\&+\left.\alpha_{LR}^{\mu u_i}\lambda_L^{\mu u_i} \lambda_R^{\tau u_i}\right)\Big\|^2+\left(L\leftrightarrow R\right),
\end{align}
where  $\alpha_{LL}^{\mu u_i}$, etc. stand for the loop integrals. To simplify our analysis we assume the  scenario where  $\lambda_R^{\ell u_i}/\lambda_L^{\ell u_i}=O(\epsilon) $  ($\ell=\mu,\,\tau$), with $\epsilon=10^{-3}$(predominantly left-handed couplings), $10^{-1}$ (small right handed-couplings), and 1 (purely scalar couplings).   We do not analyze the case when $\epsilon>1$ as a similar situation is observed as in the $\epsilon<1$ case but with $\lambda_L^{\ell u_i}$ replaced by $\lambda_R^{\ell u_i}$. Thus the parameter $\epsilon$ is a measure of the relative size between the right and left-handed LQ couplings.  Under this assumption, the $\tau\to\mu\gamma$ decay width becomes a function of the products $\lambda_L^{\mu c} \lambda_L^{\tau c}$ and $\lambda_L^{\mu t}\lambda_L^{\tau t }$.
We thus show in Fig. \ref{LFVBoundI1} the allowed area in the $\lambda_L^{\mu c} \lambda_L^{\tau c}$ vs $\lambda_L^{\mu t}\lambda_L^{\tau t }$ plane for three values of $m_{\Omega_{5/3}}$. We observe that for $m_{\Omega_{5/3}}=1000$ GeV  the largest allowed area is obtained in the scenario with $\epsilon=10^{-3}$, which allow $\lambda_L^{\mu u_i} \lambda_L^{\tau u_i}$ values as large as $O(10^{-1})$, whereas the smallest area is obtained when $\epsilon=1$, which allows $\lambda_L^{\mu u_i} \lambda_L^{\tau u_i}$ values of the order of  $O(10^{-4})$.
Such bounds are slightly relaxed when  $m_{\Omega_{5/3}}$ increases up to 2 TeV.

As far as constraints on the $| \lambda_{L,R}^{\ell u_i}|$ couplings from direct LQ searches at the LHC  via the Drell-Yan process \cite{Sirunyan:2018exx}, single production \cite{Khachatryan:2015qda}, and pair production \cite{Sirunyan:2018ryt}, an up-to-date discussion is presented in Ref.  \cite{Schmaltz:2018nls}. A restricted  scenario (minimal LQ model) is considered where each LQ is allowed to couple to just one lepton-quark pair.  In particular, a 95\% C.L. limit on  $| \lambda^{\mu c}|$ of the order of $O(1)$ is obtained for a LQ with a mass above the 1 TeV level  from  the Drell-Yan process \cite{Sirunyan:2018exx}, whereas the bounds obtained from the LHC Run 1 and Run 2  data on single production \cite{Khachatryan:2015qda} yield less stringent bounds. Although such limit could be relaxed in a more general scenario where the LQ is allowed to couple to more than one fermion pair, below  we assume a conservative scenario and consider the bound $| \lambda^{\mu c}|\le O(1)$, whereas  for the remaining couplings we impose the $|\lambda_{L,R}^{\mu t}|,\,|\lambda_{L,R}^{\tau u_i}|< 4\pi$ bound  to avoid the breakdown of perturbativity.

\begin{figure*}
\centering\includegraphics[width=14cm]{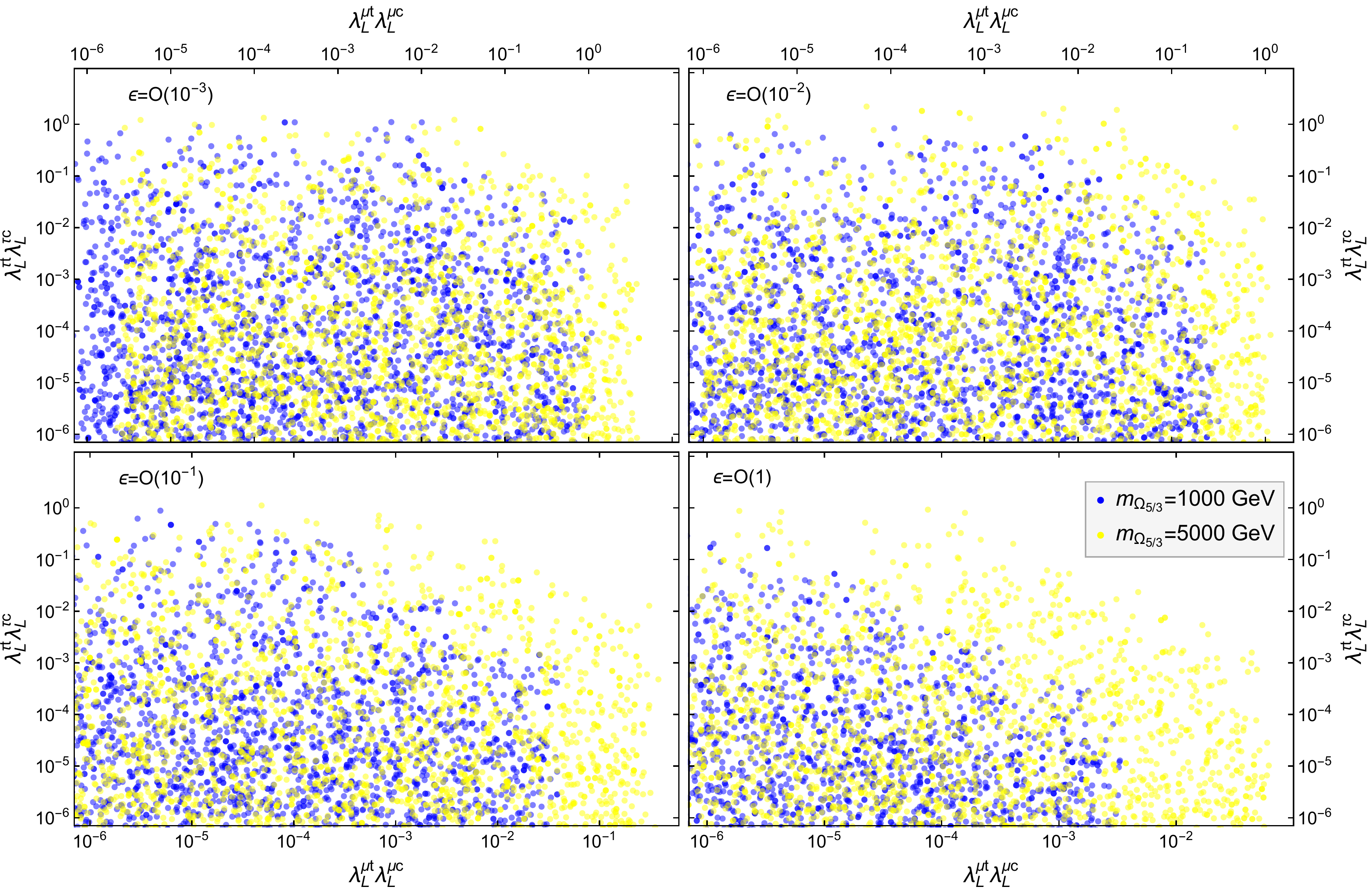}
\caption{Allowed areas with 95\% C.L. in the $\lambda_L^{\mu t}\lambda_L^{\mu c}$ vs $\lambda_L^{\tau t }\lambda_L^{\tau c}$ plane consistent with both  the $\Delta a_\mu$ discrepancy  and  the experimental bound on the $\tau\to\mu\gamma$ decay for $m_{\Omega_{5/3}}=1000$ (dark points)  and 5000 GeV  (light
points) in the scenarios where $\lambda_R^{\ell u_i}/\lambda_L^{\ell u_i}= O(\epsilon)$, for four $\epsilon$ values.  For the
$\lambda^{\mu c}$ coupling we use the constraint
$| \lambda^{\mu c}|\le O(1)$ from direct searches at the LCH \cite{Schmaltz:2018nls}, whereas for the remaining couplings we use the additional  constraint $| \lambda_{L,R}^{\ell u_i}|\le 4\pi$ to avoid  perturbativity violation. Here $\ell=\mu,\,\tau$ and $u_i=c,\,t$. \label{LamtauboundI}}
\end{figure*}

We are interested in the region of the parameter space where the largest $t\to c X$ and $t\to c\bar\ell\ell$ branching ratios can be reached, which is the area where either $\lambda_L^{\tau t} \lambda_L^{\tau c}$ or $\lambda_L^{\mu t} \lambda_L^{\mu c}$ reaches their largest allowed values.  Again we consider the scenario with $\lambda_R^{\ell u_i}/\lambda_L^{\ell u_i}=O(\epsilon)$, with four $\epsilon$ values, and perform a scan of $(\lambda_L^{\mu c},\,\lambda_L^{\mu t}, \, \lambda_L^{\tau c},\,\lambda_L^{\tau t})$ points  consistent with both the $\Delta a_\mu$ discrepancy (Fig. \ref{amuBoundI}) and  the constraint on the $\tau\to \mu\gamma$ decay (Fig. \ref{LFVBoundI1})  for two values of $m_{\Omega_{5/3}}$: we consider  a large mass splitting to observe  how  the LQ couplings get constrained by the experimental data.  As already discussed, we also impose the bound $| \lambda_{L,R}^{\mu c}|\le O(1)$ from the direct LQ search at the LHC and, to avoid  perturbativity violation, we impose  the extra constraint $| \lambda_{L,R}^{\mu t}|,\,| \lambda_{L,R}^{\tau u_i}|< 4\pi$. The corresponding allowed areas in the $\lambda_L^{\mu t } \lambda_L^{\mu c}$ vs  $\lambda_L^{\tau t }\lambda_L^{\tau c}$ plane are  shown in Fig. \ref{LamtauboundI}.  We observe that, for $m_{\Omega_{5/3}}=1000$ GeV,  the scenario with  $\epsilon=10^{-3}$ (top left plot)  allows values of $\lambda_L^{\tau c}\lambda_L^{\tau t}$ as large as $O(1)$ for  $\lambda_L^{\mu t} \lambda_L^{\mu c}$ of the order of $10^{-1}$, but values of the order of the order of $O(1)$ are allowed for $\lambda_L^{\mu c}\lambda_L^{\mu t}$  for  $\lambda_L^{tau t} \lambda_L^{\tau c}$ of the order of $10^{-2}$. For fixed $\epsilon$, the allowed area expands slightly when the LQ mass increases, which is expected as the loop functions become suppressed for large LQ mass, thereby allowing larger couplings. On the other hand, for fixed $m_{\Omega_{5/3}}$ GeV,  the allowed areas shrink significantly in the $\lambda_L^{\mu t} \lambda_L^{\mu c}$ direction and slightly  in the $\lambda_L^{\tau t} \lambda_L^{\tau c}$ direction as $\epsilon$ increases.  For instance, in the scenario when $\epsilon=1$ (bottom right plot), the largest allowed  $\lambda_L^{\mu t}\lambda_L^{\mu c}$ values for $m_{\Omega_{5/3}}=1000$ GeV  are of the order of $O(10^{-3})$ for small  $\lambda_L^{\tau t} \lambda_L^{\tau c}$, whereas the latter can be as large as $O(10^{-1})$  for very small $\lambda_L^{\mu t}\lambda_L^{\mu c}$.
We conclude that  the scenario with predominantly dominant left-handed couplings ($\epsilon=10^{-3}$) is the one that allows the largest values of the LQ couplings.

\section{Numerical analysis of the $t\to c X$ and $t\to c\bar\ell\ell$ branching ratios}
\label{NumAnalysis}
We now turn to analyze  the behavior of the $t\to c X$ and $t\to c\bar\ell\ell$ branching ratios in the allowed area of the parameter space.  For the numerical evaluation of the one-loop induced decays $t\to c X$ we have made a cross-check by evaluating the Passarino-Veltman scalar functions  via the LoopTools package \cite{vanOldenborgh:1989wn,Hahn:1998yk} and then comparing the results with those obtained by numerical integration of the parametric integrals. For the tree-level induced decay $t\to c\bar\ell\ell$ we have used the Mathematica numerical integration routines to solve the two-dimensional integral of Eq. \eqref{fitofjfmfldecay}.

\subsection{$t\to c X$ branching ratios}
We first consider two $\epsilon$ values and present in Table \ref{BRs}  a few sets of allowed $(\lambda_L^{\mu c}$, $\lambda_L^{\mu t }$, $\lambda_L^{\tau c}$, $\lambda_L^{\tau t })$ points   where  the   $t\to cX$ decays can reach their largest branching ratios for three LQ masses. In the scenario where $\epsilon=10^{-3}$ we observe that there is a small area  where all of the $t\to c X$ branching ratios  can be as large as $10^{-9}-10^{-8}$ for $m_{\Omega_{5/3}}=1000$ GeV, though they get  suppressed by one order of magnitude when $m_{\Omega_{5/3}}$ increases up to 2000 GeV. In such an area, the LQ couplings  $\lambda_L^{\mu u_i}$ are rather small, whereas the $\lambda_L^{\tau t}$ one is very close to the perturbative limit, which means that this possibility would require a large amount of fine-tuning. As for the $\epsilon =   10^{-1}$ scenario, we observe that  the $t\to cX$ branching ratios are much smaller than in the $\epsilon =   10^{-3}$ scenario: they can be of the order of $10^{-9}-10^{-10}$ at most for $m_{\Omega_{5/3}}=1000$ GeV, and decrease by one order of magnitude as $m_{\Omega_{5/3}}$ increases up to 2000 GeV. We refrain from presenting the results for the $\epsilon=1$ scenario as the $t\to cX$ branching ratios are  two orders of magnitude than in the $\epsilon=10^{-3}$ scenario.

We also observe in Table \ref{BRs}  that all the branching ratios ${\rm Br}(t\to c X)$  are of similar order of magnitude, with ${\rm Br}(t\to cZ)$ slightly larger.  It seems surprising  that ${\rm Br}(t\to c g)$ is about the same size than ${\rm Br}(t\to c \gamma)$, whereas in the SM and other of its  extensions it is one or two orders of magnitude larger. To explain this result,  let us examine the case of the SM, where the $t\to c \gamma$ decay proceeds via a Feynman diagram where the photon emerges off a down-type quark and so the squared amplitude for the analogue $t\to cg$ diagram has an enhancement factor of $c_F \left[g_S/(-1/3 e)\right]^2\simeq O(10^2)$, where $c_F=4/3$ is the color factor. On the other hand, in our LQ model  the photon emerges off the charge $5/3e$ LQ, which means that  the enhancement factor for the squared $t\to cg$ amplitude is just  $c_F \left[g_S/(5/3 e)\right]^2\simeq O(1)$. Furthermore, in our LQ model the Feynman diagram where the photon emerges off the LQ gives a smaller contribution than  that where it emerges off the lepton, which is absent in the $t\to cg$ decay. These two facts conspire to yield   ${\rm Br}(t\to c g)\gtrsim {\rm Br}(t\to c \gamma)$.  It is also worth mentioning that the $t\to c H$ decay receives its main contribution from the diagram where the Higgs boson is emitted off the LQ line, and thus its decay width is  very sensitive to the magnitude of the $\lambda_{\Omega_{5/3}}$ coupling.

\begin{table*}[!htb]
\caption{Branching ratios of the $t\to cX$  decays for a few $(\lambda_L^{\mu c}, \lambda_L^{\mu t },\lambda_L^{\tau c} \lambda_L^{\tau t })$ points inside the area allowed by   the $\Delta a_\mu$ discrepancy  and  the experimental bound on the $\tau\to\mu\gamma$ decay. We  consider the scenario where $\lambda_R^{\ell u}/\lambda_L^{\ell u_i}=O(\epsilon)$ for $\epsilon=10^{-3}$ and $10^{-1}$. For the coupling of the Higgs boson to a  LQ pair we use $\lambda_{\Omega_{5/3}}=1$.   The  $t\to cX$  branching ratios are given in units of $10^{-8}$ for $\epsilon=10^{-3}$ and $10^{-9}$  for $\epsilon=10^{-1}$.
\label{BRs}}
\centering\begin{tabular}{cCCCCCCCC}
\hline
\hline
\multicolumn{9}{c}{$\lambda_R^{\ell u}/\lambda_L^{\ell u_i}=O(10^{-3})$}\\
\hline
$m_{\Omega_{5/3}}$ [GeV]&\lambda_L^{\mu c}&\lambda_L^{\mu t}&\lambda_L^{\tau c}&\lambda_L^{\tau t}&\gamma&g&Z&H\\
\hline
 1000&2.54\times 10^{-4}&7.8\times 10^{-1}&2.39\times 10^{-1}&9.48&0.416&0.665&1.02&0.68\\
&1.48\times 10^{-4}&6.54\times 10^{-1}&2.06\times 10^{-1}&7.87&0.212&0.339&0.521&0.347\\
&2.52\times 10^{-5}&1.02&1.72\times 10^{-1}&7.86&0.147&0.236&0.362&0.241\\
&1.25\times 10^{-5}&7.68\times 10^{-1}&1.28\times 10^{-1}&9.97&0.132&0.212&0.325&0.216\\
&8.47\times 10^{-5}&6.48\times 10^{-1}&2.54\times 10^{-1}&4.97&0.129&0.207&0.318&0.211\\
\hline
 1500&1.31\times 10^{-3}&1.&3.49\times 10^{-1}&6.52&0.0819&0.131&0.246&0.135\\
&3.45\times 10^{-4}&9.21\times 10^{-1}&2.56\times 10^{-1}&8.46&0.0741&0.119&0.222&0.122\\
&1.02\times 10^{-5}&1.04&2.58\times 10^{-1}&8.04&0.0678&0.109&0.203&0.111\\
&2.17\times 10^{-6}&1.21&2.89\times 10^{-1}&6.89&0.0624&0.1&0.187&0.103\\
&1.65\times 10^{-4}&1.13&2.5\times 10^{-1}&6.84&0.0462&0.074&0.138&0.0758\\
\hline
 2000&2.16\times 10^{-5}&1.1&4.48\times 10^{-1}&7.39&0.0543&0.0872&0.186&0.0894\\
&7.69\times 10^{-6}&1.29&3.29\times 10^{-1}&9.86&0.0524&0.0841&0.179&0.0862\\
&2.59\times 10^{-3}&1.07&3.69\times 10^{-1}&8.71&0.0512&0.082&0.175&0.0843\\
&6.09\times 10^{-3}&1.07&3.05\times 10^{-1}&8.34&0.0323&0.0514&0.111&0.0532\\
&3.9\times 10^{-4}&1.69&2.73\times 10^{-1}&8.98&0.0298&0.0478&0.102&0.049\\
\hline
\multicolumn{9}{c}{$\lambda_R^{\ell u}/\lambda_L^{\ell u_i}=O(10^{-1})$}\\
\hline
$m_{\Omega_{5/3}}$ [GeV]&\lambda_L^{\mu c}&\lambda_L^{\mu t}&\lambda_L^{\tau c}&\lambda_L^{\tau t}&\gamma&g&Z&H\\
\hline
 1000&2.54\times 10^{-4}&7.8\times 10^{-1}&2.39\times 10^{-1}&9.48&4.51&6.73&10.2&6.92\\
&1.48\times 10^{-4}&6.54\times 10^{-1}&2.06\times 10^{-1}&7.87&2.3&3.43&5.21&3.53\\
&2.52\times 10^{-5}&1.02&1.72\times 10^{-1}&7.86&1.6&2.39&3.62&2.45\\
&1.25\times 10^{-5}&7.68\times 10^{-1}&1.28\times 10^{-1}&9.97&1.44&2.14&3.25&2.2\\
&8.47\times 10^{-5}&6.48\times 10^{-1}&2.54\times 10^{-1}&4.97&1.4&2.09&3.17&2.15\\
\hline
 1500&1.31\times 10^{-3}&1.&3.49\times 10^{-1}&6.52&0.893&1.33&2.45&1.37\\
&3.45\times 10^{-4}&9.21\times 10^{-1}&2.56\times 10^{-1}&8.46&0.809&1.2&2.22&1.24\\
&1.02\times 10^{-5}&1.04&2.58\times 10^{-1}&8.04&0.739&1.1&2.03&1.13\\
&2.17\times 10^{-6}&1.21&2.89\times 10^{-1}&6.89&0.681&1.01&1.87&1.04\\
&1.65\times 10^{-4}&1.13&2.5\times 10^{-1}&6.84&0.504&0.749&1.38&0.772\\
\hline
 2000&2.16\times 10^{-5}&1.1&4.48\times 10^{-1}&7.39&0.595&0.883&1.86&0.911\\
&7.69\times 10^{-6}&1.29&3.29\times 10^{-1}&9.86&0.574&0.851&1.79&0.878\\
&2.59\times 10^{-3}&1.07&3.69\times 10^{-1}&8.71&0.561&0.83&1.75&0.859\\
&6.09\times 10^{-3}&1.07&3.05\times 10^{-1}&8.34&0.354&0.52&1.11&0.542\\
&3.9\times 10^{-4}&1.69&2.73\times 10^{-1}&8.98&0.326&0.484&1.02&0.5\\
\hline
\hline
\end{tabular}
 \end{table*}

\begin{figure*}[!htb]
\centering\includegraphics[width=14cm]{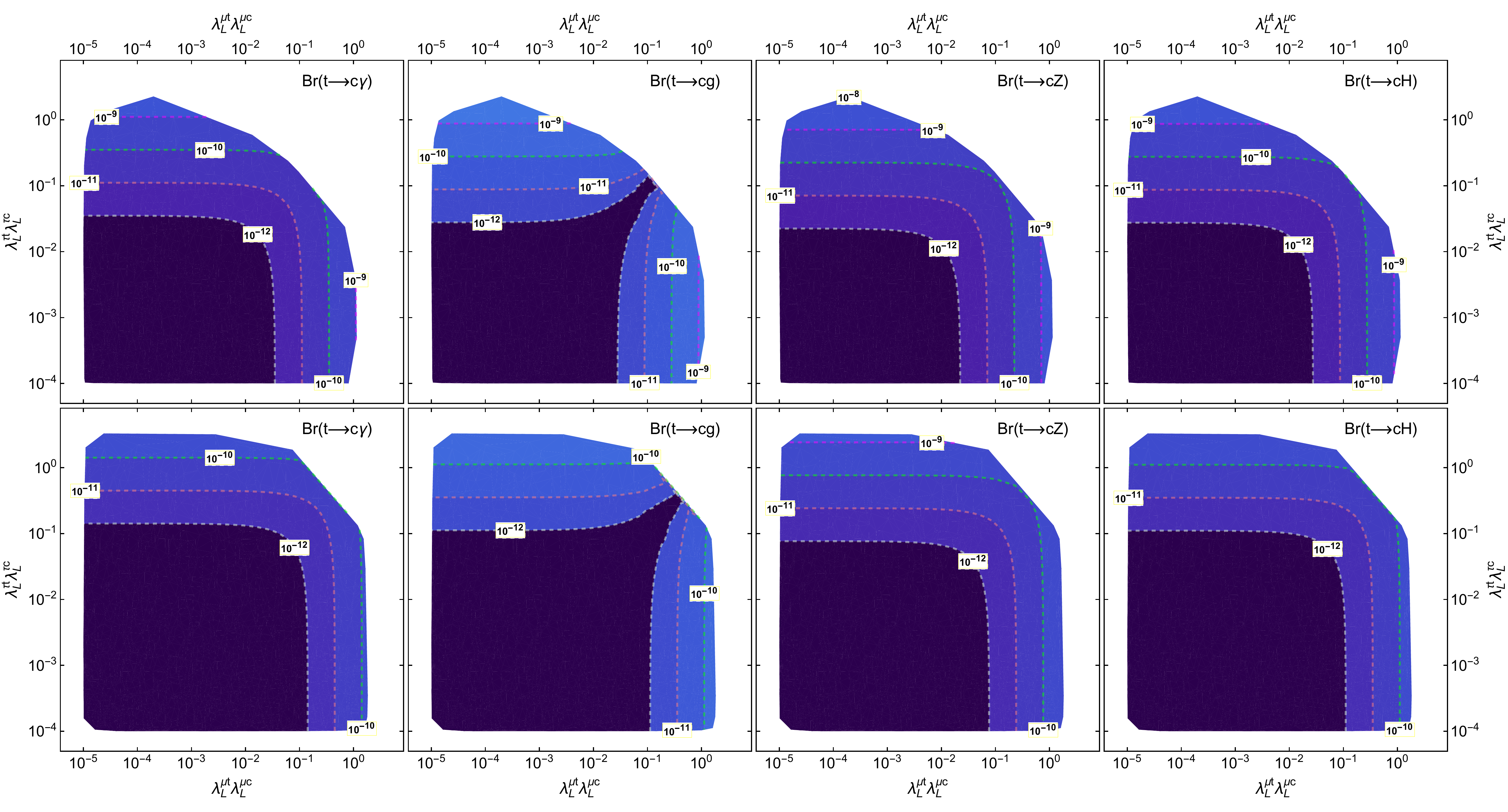}
\caption{Contours of the branching ratios of the $t\to c X$ decays for $m_{\Omega_{5/3}}=1000$ (top plots) and  2000 GeV (bottom plots) in the $\lambda_L^{\mu t}\lambda_L^{\mu c}$ vs $\lambda_L^{\tau t }\lambda_L^{\tau c}$ plane for $(\lambda_L^{\mu c}, \lambda_L^{\mu t },\lambda_L^{\tau c} \lambda_L^{\tau t })$ values inside the area allowed by   the $\Delta a_\mu$ discrepancy  and  the experimental bound on the $\tau\to\mu\gamma$ in the scenario with $\lambda_R^{\ell u}/\lambda_L^{\ell u_i}=O(10^{-3})$. \label{contourplotI}}
\end{figure*}

 Finally we show in Fig.  \ref{contourplotI}   the contours of the $t\to c X$   branching ratios in the allowed area of the $\lambda_L^{\mu t}\lambda_L^{\mu c}$ vs $\lambda_L^{\tau t }\lambda_L^{\tau c}$ plane  in the scenario with $\lambda_R^{\ell u_i}/\lambda_L^{\ell u_i}=O(10^{-3})$, where the largest values of the $t\to c X$ branching ratios are reached.   As already noted, when  $m_{\Omega_{5/3}}=1000$ GeV the largest   $t\to c X$ branching ratios, of the order of $10^{-9}-10^{-8}$, are  obtained in a tiny area where $\lambda_L^{\mu t}\lambda_L^{\mu c}$ is very small and $\lambda_L^{\tau t }\lambda_L^{\tau c}$ reaches its largest allowed values (top-left corner of the upper plots), but they  decrease as the allowed area expands.  It means that the largest branching ratios are obtained in the region where  the main contribution arises from the loops with an internal tau lepton, which is due to the fact that the LQ couplings to the tau lepton are less constrained than those to the muon. We also observe that the $t\to c X$  branching ratios   decrease by  one or two orders of magnitude as  $m_{\Omega_{5/3}}$ reaches the 2 TeV level, where  they  can be as large as $10^{-9}-10^{-10}$. The behavior of the $t\to cX$ branching ratios in the  scenarios with $\epsilon=10^{-1}$ and $\epsilon=1$ is rather  similar to that observed in Fig. \ref{contourplotI}, but they are one or two orders of magnitude below:  they  can only be as large as $10^{-9}$ for $\epsilon=10^{-1}$ and $10^{-10}$  for $\epsilon=1$.

\subsection{$t\to c\bar\ell\ell$ branching ratios}
\begin{figure*}[!htb]
\centering\includegraphics[width=12cm]{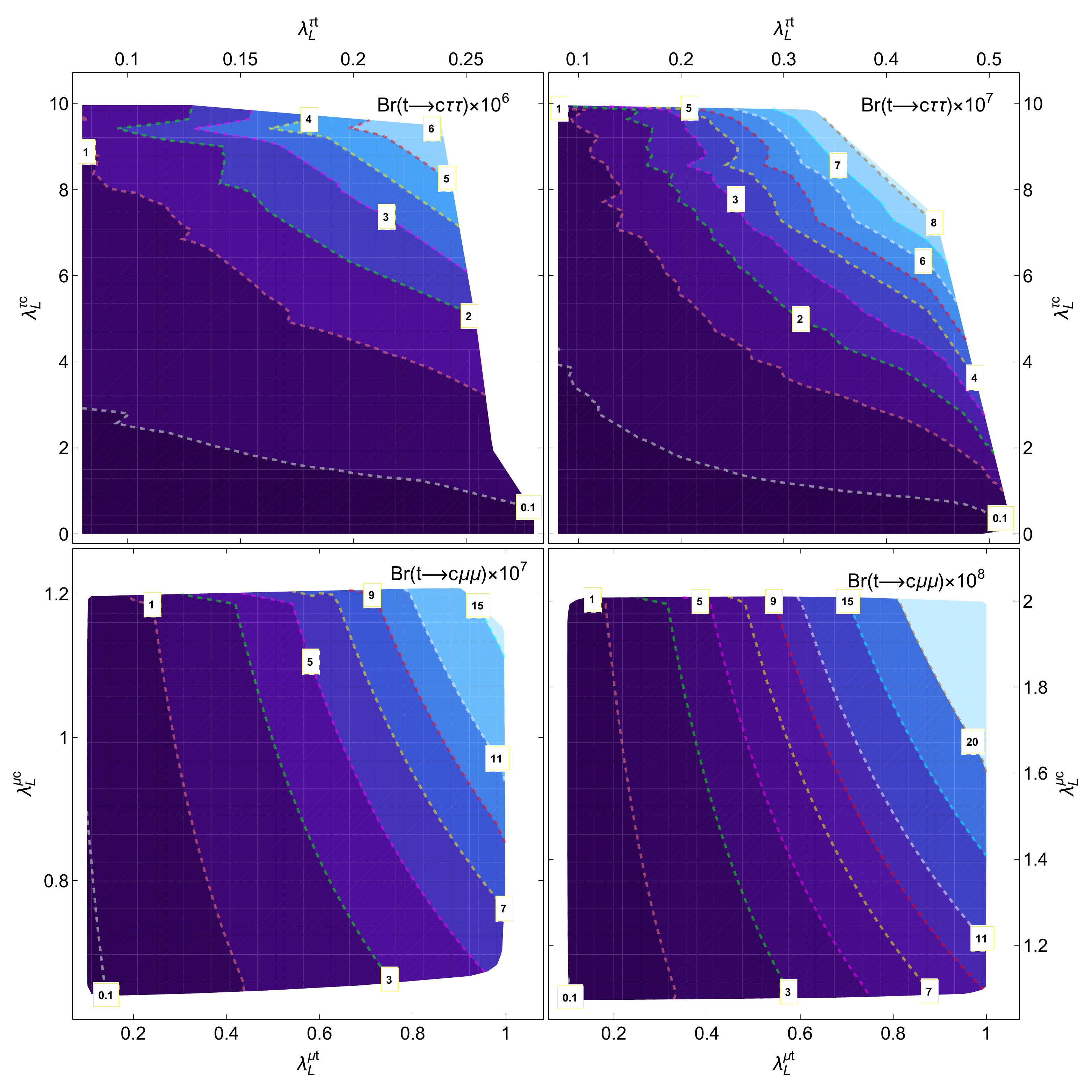}
\caption{Contours of the branching ratios of the $t\to c \bar\tau\tau$ and $t\to c \bar\mu\mu$  decays for $m_{\Omega_{5/3}}=1000$ GeV  (left plots) and  2000 GeV (right plots) in the $\lambda_L^{\ell t}$ vs $\lambda_L^{\ell c}$ plane for $(\lambda_L^{\mu c}, \lambda_L^{\mu t },\lambda_L^{\tau c} \lambda_L^{\tau t })$ values inside the area allowed by   the $\Delta a_\mu$ discrepancy  and  the experimental bound on the $\tau\to\mu\gamma$ in the scenario with $\lambda_R^{\ell u}/\lambda_L^{\ell u_i}=O(10^{-3})$.\label{contourplotIII}}
\end{figure*}

We now perform the corresponding analysis for the $t\to c\bar\ell\ell$ ($\ell=\mu,\,\tau$)  branching ratios in  the area allowed by the experimental constraints discussed above. In Fig. \ref{contourplotIII} we show the contours of ${\rm Br}(t\to c\bar\ell\ell)$ in the  $\lambda_L^{\ell t}$ vs $\lambda_L^{\ell c}$ plane  in the scenario with $\lambda_R^{\ell u_i}/\lambda_L^{\ell u_i}=O(10^{-3})$ for two values of the LQ mass. We observe that for $m_{\Omega_{5/3}}=1000$ GeV, ${\rm Br}(t\to c\bar\tau\tau)$ can be of as large as $10^{-6}$, whereas ${\rm Br}(t\to c\bar\mu\mu)$ is one order of magnitude below, which is due to the fact that the $\lambda_L^{\mu q}$ couplings are more constrained than the  $\lambda_L^{\tau q}$ ones. When   $m_{\Omega_{5/3}}$ increases up to 2000 GeV, the $t\to c\bar\ell\ell$  branching ratios decrease by about one order of magnitude. As for the $t\to c\bar\mu\tau$ decay, its  branching ratio is also suppressed as  involves the $\lambda_L^{\mu q}$ couplings. In conclusion, the three-body tree-level decay $t\to c\bar\ell\ell$ can have larger branching ratios than the two-body one-loop decays $t\to cX$.

\section{Summary and outlook}
\label{Conclusions}
The FCNC decays of the top quark $t\to cX$ ($X=\gamma,\,g,\,Z,\,H$) and $t\to c\bar\ell\ell$ ($\ell=\mu,\,\tau$) were calculated in a simple LQ model with no proton decay, where the SM is augmented by a $SU(2)$ scalar LQ doublet with hypercharge  $Y=7/6$. In such a model there is  a non-chiral LQ with electric charge $Q=5/3e$ that couples to charged leptons and up quarks and contribute to the FCNC decays of the top quark.

As far as the analytical results are concerned, we perform a general calculation  of the FCNC fermion decays $f_i\to  f_j X$ and $f_i\to f_j\bar f_m f_l$. The loop amplitudes of the  $f_i\to  f_j X$ decays are presented in terms of both Passarino-Veltman scalar functions and Feynman parameter integrals, which can be  useful to calculate the contributions of other scalar LQs. On the other hand, an analytical expression is presented for the $f_i\to f_j\bar f_m f_l$  decay width, which can be numerically evaluated.

As for the numerical analysis, to obtain bounds on the parameter space of the model we assumed that the LQ only couples to the fermions of the last two families and  used the experimental constraints on the LHC Higgs boson data, the muon anomalous magnetic dipole moment $a_\mu$,  the LFV decay of the tau lepton $\tau\to \mu\gamma$,  as well as the direct LQ searches at the LHC via the Drell-Yan process, single production, and double production. For the LQ couplings to charged leptons and up quarks $\lambda_{L,R}^{\ell u_i}$, a scenario was considered where $\lambda_R^{\ell u}/\lambda_L^{\ell u_i}=O(\epsilon)$, with $\epsilon$ being a measure of the relative size between the right- and left-handed LQ couplings. Afterwards,  the $t\to cX$ and $t\to c\bar\ell\ell$ branching ratios were evaluated in the allowed region of the parameter space. In particular, we find that in the scenario where the LQ couplings  are  predominantly left-handed, $\epsilon=O(10^{-3})$, there is a tiny region of the parameter space where the branching ratios of the one-loop induced $t\to cX$ decay can be as large as $10^{-8}$ for $m_{\Omega_{5/3}}=1000$ GeV, with the main contribution arising from the loops with an internal tau lepton, although a large amount of fine-tuning between the LQ couplings would be required. However, for $\epsilon=10^{-1}$ ($\epsilon=1$), the main part of the allowed region yields   $t\to cX$ branching ratios of the order of $10^{-9}$ ($10^{-10}$) at most. For $m_{\Omega_{5/3}}\ge 2000$ GeV, the largest $t\to cX$ branching ratios are of the order of $10^{-10}$ in all the scenarios analyzed in this work. Although the $t\to cX$ branching ratios are larger in our LQ model than in the SM, such contributions would be out of the reach of detection in the near future. As for the tree-level induced decays $t\to c\bar\ell\ell$, the $t\to c\bar\tau\tau$ branching ratio can be as large as $10^{-6}$ for $m_{\Omega_{5/3}}=1000$ GeV in the scenario with $\epsilon=O(10^{-3})$, but ${\rm Br}(t\to c\bar\mu\mu)$ is one order of magnitude below. These branching ratios  decrease by about one order of magnitude when the LQ mass increases up  to 2000 GeV.

It is worth noting that experimental constraints on the LQ mass and couplings obtained from the direct  search at the LHC are very stringent, but they rely on several assumptions and may be relaxed, which would yield a slight enhancement of the LQ contribution to the top quark  FCNC top quark decays.  The magnitude of the $t\to cX$ branching ratios is similar  to that recently found for the contributions   from a scalar LQ with charge $-1/3e$, which arises in a model with a scalar LQ singlet \cite{Kim:2018oih}.   We do not consider this scenario in our analysis as we are interested in LQ models where no further symmetries must be invoked to forbid the proton decay \cite{Arnold:2013cva}.

\begin{acknowledgements}We acknowledge support from Consejo Nacional de Ciencia y Tecnolog\'ia and Sistema Nacional de Investigadores. Partial support from Vicerrector\'ia de Investigaci\'on y Estudios de Posgrado de la Ben\'emerita Universidad Aut\'onoma de Puebla is also acknowledged.
\end{acknowledgements}

\onecolumn
\appendix

\section{Loop integrals for the $f_i \to f_j  V$ decay}
\label{fitofjVCoefficients}
We now present the contribution of the scalar LQ $S$ to the  $f_i \to f_j  V$ loop amplitudes. Although in these Appendices $S$ will  stand for the $\Omega_{5/3}$ LQ, as already explained, our results are also valid for the contribution of any other  scalar LQ.

\subsection{Passarino-Veltman results}
We first define the following sets of ultraviolet finite Passarino-Veltman scalar function combinations

\begin{align}
\Delta_0&=B_0(m_Z^2,m_k^2,m_k^2)-B_0(m_j^2,m_k^2,m_S^2),\\
\Delta_1&=B_0(m_Z^2,m_k^2,m_k^2)- B_0(0,m_k^2,m_S^2),\\
\Delta_2&=B_0(m_i^2,m_k^2,m_S^2) -B_0(m_j^2,m_k^2,m_S^2),\\
\Delta_3&= B_0(m_j^2,m_k^2,m_S^2) -B_0(0,m_k^2,m_S^2),\\
\Delta_4&=B_0(0,m_k^2,m_k^2) -B_0(0,m_S^2,m_S^2),\\
\Delta_5&= B_0(0,m_S^2,m_S^2) -B_0(0,m_k^2,m_S^2),\\
C_{kSk}&=m_S^2C_0(m_i^2, m_j^2, 0, m_k^2, m_S^2,m_k^2),\\
C_{SkS}&=m_S^2C_0(m_i^2, m_j^2, 0, m_S^2, m_k^2,m_S^2),\\
C_{ZkSk}&=m_S^2C_0(m_i^2, m_j^2, m_Z^2, m_k^2, m_S^2,m_k^2),\\
C_{ZSkS}&=m_S^2C_0(m_i^2, m_j^2, m_Z^2, m_S^2, m_k^2,m_S^2).
\end{align}

The form factors of Eq. \eqref{MfitofjV} are given in terms of these ultraviolet-finite  functions as follows.

\subsubsection{$f_i \to f_j  \gamma$   decay}
\label{fitofjgammaCoefficientsPV}
There are only dipole form factors as the monopole ones must vanish due to electromagnetic gauge invariance. Although each Feynman diagram has ultraviolet divergences, they cancel out when summing over all the contributions. The results read

\begin{align}
\label{LgammaPV}
L^\gamma&=\frac{N_c e}{16\pi^2}\frac{\sqrt{x_i}\xi_{ij} }{8}\Bigg(\frac{\lambda _L^{{ik}} \lambda _L^{{kj}}}{\sqrt{x_i}} \Big(x_i(Q_S-Q_k)+\frac{1}{\xi _{ij}}
\left(Q_S \left(x_i \left(x_j-2 \xi_{k}\right)+x_j \xi_{k}\right)-Q_k \left(x_i \left(x_j+2\xi_{k}\right)-x_j\xi_{k}\right)\right)
\Delta _2\nonumber\\&- \xi_{k} \left(Q_k+Q_S\right)\Delta _3-2 x_i \left(Q_k x_k C_{{kSk}}- Q_S C_{{SkS}}\right)\Big)+\frac{\lambda _R^{{ik}} \lambda _R^{{kj}}}{\sqrt{x_j}}\Big( \frac{ x_j }{\xi _{ij}}\left(Q_k \left(\xi_{k}+x_i\right)+Q_S \left(\xi_{k}-x_i\right)\right)
\Delta _2\nonumber\\&
+ 2 \xi_{k} \left(Q_k+Q_S\right)\Delta _3 +2  x_j \left(Q_k x_k C_{{kSk}}- Q_S
C_{{SkS}}\right)+x_j(Q_k-Q_S)\Big)-\frac{\sqrt{x_k} \lambda _L^{{kj}} \lambda _R^{{ik}}}{\xi_{ij}^3 }\Big(\xi _{{ij}} \left(Q_k+Q_S\right)
\Delta _2\nonumber\\&+Q_k \left(x_j^2-x_i \left(\eta _{{jk}}+\xi _{{jk}}\right)+x_i^2\right)C_{{kSk}}\Big)\Bigg),
\end{align}
where $N_c$ is the color number of the internal fermion and we introduced the following definitions $x_a=m_a^2/m_S^2$, $\xi_{ab}=x_a-x_b$, $\eta_{ab}=x_a+x_b$, $\xi_a=x_a-1$, and
$\eta_a=x_a+1$. In addition, the right handed form factors can be obtained from the left-handed ones as follows
\begin{equation}
\label{Rgamma}
R^\gamma=L^\gamma\left(\lambda^{lm}_L\leftrightarrow \lambda^{lm}_R, Q_S\to -Q_S\right).
\end{equation}

\subsubsection{$f_i \to f_j  Z$ decay}
The amplitude for this decay contains both dipole and monopole form factors. Again the ultraviolet divergences cancel when summing over partial contributions. The $L^Z$ and $L^{'Z}$ form factors are too lengthy and can be written as a sum of partial terms arising from each contributing diagram as follows

\begin{equation}
\label{LZ}
L^Z=\frac{N_cg}{32\pi^2c_W}\frac{\sqrt{x_i}}{8}\sum_{j={\rm a,b}}\sum_{i=1}^3L^{\rm(j)}_i,
\end{equation}
and
\begin{equation}
\label{LZ'}
L^{'Z}=\frac{N_cg }{32\pi^2 c_W}\frac{\sqrt{x_i}}{8}\sum_{j={\rm a,b,cd}}\sum_{i=1}^3 L^{'\rm (j)}_{i},
\end{equation}
where the superscript stands for the Feynman diagram of Fig. \ref{fitofjVdecay} out of which the corresponding term arises, with (cd) standing for the sum of the contributions of diagrams (c) and (d).

The contributions of diagram (a) are given by

\begin{align}
L_1^{(\rm a)}&=\frac{\sqrt{x_j} g_R^k \lambda _R^{{ik}} \lambda _R^{{kj}}}{2\delta ^2 }\Bigg(\Big(x_Z^3
+x_Z^2 \left(4 x_i-2 x_j-6 {\xi_{k}}\right)
-\xi _{{ij}} x_Z \left(x_j+6 {\xi_{k}}+5 x_i\right)
\Big)\Delta _1\nonumber\\
&+\Big(x_Z^2\left(\xi_{k}-5 x_i\right)
+2 x_Z \left(2x_i\eta_{ij} +5 x_i \xi_{k}-x_j \xi_{k}\right)
+\xi _{{ij}}^2 \left(\eta _{{ik}}-1\right)
\Big)\Delta _2\nonumber\\
&+\frac{1}{x_j}\Big(x_Z^3\left(1-\eta _{{jk}}\right)
+x_Z^2 \left(x_i \left(3 {\xi_{k}}-4x_j\right)+x_j \left(2 x_j+7 {\xi_{k}}\right)\right)
\nonumber\\&
+\xi _{{ij}} x_Z \left(x_i \left(5 x_j-3 {\xi_{k}}\right)+x_j \left(x_j+5 {\xi_{k}}\right)\right)
+\xi _{{ij}}^3 {\xi_{k}}
\Big)\Delta _3\nonumber\\
&+2\Big(x_Z^3\left(\xi _{{ik}}+2\right)
+x_Z^2 \left(x_i \left(x_j-3 x_k+2\right)+x_j \left(x_k-4\right)+3 {\xi_{k}}^2-2 x_i^2\right)\nonumber\\
&+\xi _{{ij}} x_Z \left(x_i \left(2 x_j+3 x_k-4\right)-x_j \left(x_k+2\right)+3 {\xi_{k}}^2+x_i^2\right)
+\xi _{{ij}}^3 x_k
\Big)C_{ZkSk} +\delta \left(\xi _{{ij}}+x_Z\right)\Bigg),
\end{align}

\begin{align}
L_2^{(\rm a)}&=\frac{g_L^k \lambda _L^{{ik}} \lambda _L^{{kj}}}{2\sqrt{x_i}\delta ^2 }\Bigg(x_i\Big(x_Z^3
-2 x_Z^2 \left(3\xi_{k}-2 x_j+x_i\right)
+\xi _{{ij}} x_Z \left(5 x_j+6 \xi_{k}+x_i\right)
\Big)\Delta _1\nonumber\\
&+\Big(x_Z^3\left(1-\eta _{{ik}}\right) \
+\left(x_j+2 x_i\right) \left(3 \xi_{k}+x_i\right) x_Z^2
-x_Z \left(x_i^2 \left(8 x_j+3 \xi_{k}\right)\right.\nonumber\\
&\left.-x_i x_j \left(x_j-2 \xi_{k}\right)+3 x_j^2 \xi_{k}+x_i^3\right)
-\xi _{{ij}}^2 \left(x_i \left(x_j+2 \xi_{k}\right)-x_j\xi_{k}\right)
\Big)\Delta _2\nonumber\\
&+\Big( x_Z^3\left(1-\eta _{{ik}}\right)
+x_Z^2 \left(x_i \left(7 \xi_{k}-4 x_j\right)+3 x_j \xi_{k}+2 x_i^2\right)
\nonumber\\
&-\xi _{{ij}} x_Z \left(x_i \left(5 \eta _{{jk}}+x_i-5\right)-3 x_j \xi_{k}\right)
-\xi _{{ij}}^3 \xi_{k}
\Big)\Delta _3\nonumber\\
&+2x_i\Big(x_Z^3\left(\xi _{{jk}}+2\right)
+x_Z^2 \left(x_j \left(2-3 x_k\right)-2 x_j^2+x_i \left(\eta _{{jk}}-4\right)+3 \xi_{k}^2\right)
\nonumber\\
&-\xi _{{ij}} x_Z \left(\left(3 x_j-x_i-6\right) x_k+\xi_{j} \left(x_j+2 x_i-3\right)+3 x_k^2\right)
-\xi _{{ij}}^3 x_k
\Big)C_{ZkSk}+\delta x_i\left(x_Z-\xi _{{ij}}\right)\Bigg),
\end{align}

\begin{align}
L_{3}^{{\rm (a)}}&=\frac{\sqrt{x_k}  \lambda _L^{{kj}} \lambda _R^{{ik}}}{\delta }\Bigg(g_L^k\bigg(
\Big(x_Z-\xi _{{ij}}\Big)\Delta _0+\Big(\eta _{{ij}}-x_Z\Big)\Delta _2+\Big(\left(\xi _{{jk}}+1\right) x_Z+\xi _{{ij}} \left(\eta _{{jk}}-1\right)\Big)C_{{kSk}}\bigg)
\nonumber\\&+ g_R^k\Bigg(
\Big(x_Z+\xi _{{ij}}\Big)\Delta _0-2 x_i\Delta _2+\Big(x_Z \left(x_i-\xi _k\right)
+x_j \xi _k+x_i \left(\xi _{{jk}}-\xi_i\right)
\Big)C_{{kSk}}\Bigg),
\end{align}

\begin{align}
L_{1}^{'(\rm a)}&=\frac{ g_R^k \lambda _R^{{ik}}\lambda_R^{kj}}{2\delta ^2}\Bigg(\Big(-x_Z^4
+2x_Z^3 \left(2 \eta_{ij}- \xi_{k}\right)
-x_Z^2 \left(4x_i x_j+2\eta_{ij} \xi_{k}+5 \xi_{ij}^2\right)
+2 \xi _{{ij}}^2 x_Z \left(\eta _{{ij}}+2 \xi_{k}\right)
\Big)\Delta _1
\nonumber\\&+\Big(x_Z^2 \left(x_j\xi_{k}-x_i \left(5 (\eta_{j}-x_k)+x_i\right)\right)
+2 \xi _{{ij}} x_Z \left(x_j \xi_{k}+x_i \left(2(1- \eta _{{jk}})+x_i\right)\right)
-\xi _{{ij}}^3 \left(\eta _{{ik}}-1\right)\Big)\Delta _2
\nonumber\\&+\Big(x_Z^2 \left(2 x_i \left(3\xi_{k}-5 x_j\right)-x_j \left(x_j-6 \xi_{k}\right)-x_i^2\right)
+2 \xi _{{ij}}^2 x_Z \left(\eta _{{ij}}-3 \xi_{k}\right)
-\xi _{{ij}}^4
\Big)\Delta _3\nonumber\\
&+2\Big(x_Z^3 \left(2 \eta_{ij}-\eta _{{ij}} x_k+x_i x_j+\xi_{k}^2\right)
+x_Z^2 \left(x_i \left(x_j \left(4-6 x_k\right)+x_j^2+\xi_{k}^2\right)+x_j \left(x_j \left(x_k-4\right)+\xi_{k}^2\right)
\right.\nonumber\\&
\left.+x_i^2 \left(\eta _{{jk}}-4\right)\right)
+\xi _{{ij}}^2 x_Z \left(x_j \left(x_k+2\right)+x_i \left(x_k-2 \xi_{j}\right)-2 \xi_{k}^2\right)
-\xi _{{ij}}^4 x_k
\Big)C_{ZkSk}+\delta  x_Z \left(x_Z-\eta _{{ij}}\right)-\delta ^2 B_{{kS}}\Bigg)\nonumber\\&
+g_L^k \lambda _R^{{ik}} \lambda _R^{{kj}} x_k C_{ZkSk},
\end{align}

\begin{align}
L_2^{'(\rm a)}&=\frac{g_L^k \lambda _L^{{ik}} \lambda _L^{{kj}}}{2\sqrt{x_i} \sqrt{x_j} \delta ^2}\Bigg(2x_i x_j \delta x_Z+6x_i x_j \Big( x_Z^3
-x_Z^2 \left(\eta _{{ij}}+2\xi_{k}\right)
\Big)\Delta _1-x_j\Big(x_Z^3\left(\xi_{k}+3 x_i\right)
\nonumber\\&
+x_Z^2 \left(x_i \left(7(1- \eta _{{jk}})+x_i\right)-3 x_j \xi_{k}\right)
-\xi _{{ij}} x_Z \left(x_i \left(5 x_j+7\xi_{k}\right)+3 x_j \xi_{k}+5 x_i^2\right)
-\xi _{{ij}}^3 \left(1-\eta _{{ik}}\right)
\Big)\Delta _2
\nonumber\\&
-\Big(x_Z^3 \left(\eta _{{ij}} \xi_{k}+6 x_i x_j\right)
-x_Z^2 \left(3 x_i^2 \left(2 x_j+\xi_{k}\right)+2 x_i x_j \left(3 x_j+7 \xi_{k}\right)+3 x_j^2 \xi_{k}\right)
+3 \eta _{{ij}} \xi _{{ij}}^2 \xi_{k} x_Z
-\xi _{{ij}}^4 \xi_{k}
\Big)\Delta _3\nonumber\\
&+2x_i x_j\Big(x_Z^4
-2 x_Z^3 \left(\eta _{{ij}}+2 x_k-3\right)
+x_Z^2 \left(2 x_i x_j+2 \eta_{ij} \left(x_k-3\right)+\eta_{ij}^2+6 \xi_{k}^2\right)
+2 \xi _{{ij}}^2 x_k x_Z
\Big)C_{ZkSk}\Bigg),
\end{align}

\begin{align}
L_3^{'(\rm a)}&=\frac{ \sqrt{x_k} g_L^k}{\delta } \left(\sqrt{x_i} \lambda _L^{{ik}} \lambda _R^{{kj}}+\sqrt{x_j} \lambda _L^{{kj}} \lambda _R^{{ik}}\right)\Bigg(2 x_Z\Delta_0-\Big(x_Z
+\xi _{{ij}}\Big)\Delta _2 +\Big(x_Z^2-x_Z \left(\eta _{{ij}}+2 \xi_{k}\right)\Big)C_{ZkSk}\Bigg),
\end{align}
where  $\delta=x_i^2-2 (x_j+x_Z) x_i+(x_j-x_Z)^2$.
The contributions of diagram (b) are
\begin{align}
L_1^{(\rm b)}&=\frac{g_{{VSS}} \sqrt{x_j} \lambda _R^{{ik}} \lambda _R^{{kj}}}{2\delta ^2 }\Bigg(\Big(-x_Z^3
+2 x_Z^2 \left(x_j-3 \xi_{k}-2 x_i\right)
+\xi _{{ij}} x_Z \left(x_j-6 \xi_{k}+5 x_i\right)
\Big)\Delta _1
\nonumber\\
&+\Big(x_Z^2\left(\xi_{k}+5 x_i \right)
-2 x_Z \left(x_i \left(2 \eta_{ij}-5 \xi_{k}\right)+x_j \xi_{k}\right)
-\xi _{{ij}}^2 \left(\xi _{{ik}}+1\right)
\Big)\Delta _2
\nonumber\\
&+\frac{1}{x_j}\Big(x_Z^3\left(\xi _{{jk}}+1\right)
+x_Z^2 \left(x_i \left(4 x_j+3 \xi_{k}\right)+x_j \left(7\xi_{k}-2 x_j\right)\right)
\nonumber\\&
-\xi _{{ij}} x_Z\left(x_j \left(x_j-5\xi_{k}\right)+x_i \left(5 x_j+3 \xi_{k}\right)\right)
+\xi _{{ij}}^3\xi_{k}
\Big)\Delta _3
-2\Big( x_Z^3\left(\xi_{k}+\eta_{ik}\right)\nonumber\\&
+x_Z^2 \left(x_i \left(\xi_j+2 (\xi_k-x_i)\right)-x_j \left(4 x_k-1\right)+3 \xi_{k}^2\right)
\nonumber\\&
-\xi _{{ij}} x_Z \left(2 \left(x_j+2 x_i\right) x_k+x_j-x_i \left(2 x_j+x_i+3\right)-3 \xi_{k}^2\right)
+\xi _{{ij}}^3
\Big)C_{ZSkS}-\delta \left(\xi _{{ij}}+x_Z\right)\Bigg),
\end{align}

\begin{align}
L_2^{(\rm b)}&=\frac{ g_{{VSS}} \lambda _L^{{ik}} \lambda _L^{{kj}}}{2 \sqrt{x_i}\delta ^2 }\Bigg(x_i\Big(-x_Z^3
+2 x_Z^2 \left(x_i-2 x_j-3 {\xi_{k}}\right)
+\xi _{{ij}} x_Z \left(6 {\xi_{k}}-5 x_j-x_i\right)
\Big)\Delta _1\nonumber\\
&+\Big( x_Z^3\left(\xi _{{ik}}+1\right)
+x_Z^2\left(x_j+2 x_i\right) \left(3 {\xi_{k}}-x_i\right)
+x_Z \left(x_i^2 \left(8 x_j-3 {\xi_{k}}\right)-x_i x_j \left(x_j+2 {\xi_{k}}\right)-3 x_j^2 {\xi_{k}}+x_i^3\right)\nonumber\\&
+\xi _{{ij}}^2 \left(x_i \left(x_j-2 {\xi_{k}}\right)+x_j {\xi_{k}}\right)
\Big)\Delta _2+\Big(x_Z^3\left(\xi _{{ik}}+1\right)+x_Z^2 \left(x_i \left(4 x_j+7 {\xi_{k}}\right)+3 x_j {\xi_{k}}-2 x_i^2\right)
\nonumber\\
&
+\xi _{{ij}} x_Z \left(3 x_j {\xi_{k}}+x_i \left(5 \xi _{{jk}}+x_i+5\right)\right)
-\xi _{{ij}}^3 {\xi_{k}}
\Big)\Delta _3-2x_i\Big(x_Z^3 \left(\xi_{j}+2 x_k\right)\nonumber\\&
-x_Z^2 \left(x_j \left(3-2 x_k\right)-x_i \left(\eta_{j}-4 x_k\right)+2 x_j^2-3 {\xi_{k}}^2\right)\nonumber\\
&+x_Z \xi _{{ij}} \left(3-2 \left(2 x_j+x_i+3\right) x_k+x_j(x_j+2 x_i +3)+3 x_k^2-x_i\right)
+\xi _{{ij}}^3\Big)C_{ZSkS}+x_i\delta\left(\xi _{{ij}}-x_Z\right)\Bigg),
\end{align}

\begin{align}
L_3^{(\rm b)}&=\frac{g_{{VSS}} \sqrt{x_k} \lambda _L^{{kj}} \lambda _R^{{ik}}}{\delta  }\Bigg(2 x_Z\Delta _0-\Big(x_Z
+\xi _{{ij}}\Big)\Delta _2 +\Big(x_Z^2
+x_Z \left(2 \xi_{k}-\eta _{{ij}}\right)
\Big)C_{ZSkS}\Bigg),
\end{align}

\begin{align}
L_1^{'(\rm b)}&=\frac{g_{{VSS}} \lambda _R^{{ik}} \lambda _R^{{kj}}}{2 \delta ^2}\Bigg(\Big(-x_Z^4
+2 x_Z^3 \left(\eta _{{ij}}-\xi_{k}\right)
-x_Z^2 \left(2 x_i \left(5 x_j+\xi_{k}\right)+x_j \left(x_j+2\xi_{k}\right)+x_i^2\right)
+4 \xi _{{ij}}^2 \xi_{k} x_Z
\Big)\Delta _1\nonumber\\
&+\Big(2 x_i x_Z^3
+x_Z^2 \left(x_i \left(3 x_j+5 \xi_{k}\right)+x_j \xi_{k}-5 x_i^2\right)
+2 \xi _{{ij}} x_Z \left(x_i \left(3 x_j-2 \xi_{k}\right)+x_j \xi_{k}+2 x_i^2\right)
+\xi _{{ij}}^3 \left(1-\eta _{{ik}}\right)
\Big)\Delta _2\nonumber\\
&+\Big(2 \eta _{{ij}} x_Z^3
-x_Z^2 \left(x_j \left(5 x_j-6 \xi_{k}\right)+x_i \left(6-6 \eta _{{jk}}+5 x_i\right)\right)
+2 \xi _{{ij}}^2 x_Z \left(2 \eta_{ij}-3 \xi_{k}\right)
-\xi _{{ij}}^4
\Big)\Delta _3\nonumber\\
&+\Big(-2 x_k x_Z^4
+x_Z^3 \left(2 x_k \left(\eta _{{ij}}-x_k+2\right)+x_i \left(4-6 x_j\right)+4 x_j-2\right)
+2 x_Z^2 \left(x_i^2 \left(3 x_j+x_k-4\right)\right.
\nonumber\\&\left.+x_i \left(x_j \left(4-6 x_k\right)+3 x_j^2-\xi_{k}^2\right)+x_j \left(x_j \left(x_k-4\right)-\xi_{k}^2\right)\right)
+2 \xi _{{ij}}^2 x_Z \left(2 \xi_{k}^2-\eta _{{ij}} \left(x_k-2\right)\right)
\Big)C_{ZSkS}\nonumber\\&+\delta  x_Z \left(\eta _{{ij}}-x_Z\right)-\delta ^2 B_{{kS}}
\Bigg),
\end{align}

\begin{align}
L_2^{'(\rm b)}&=\frac{g_{{VSS}}\sqrt{x_i} \sqrt{x_j} \lambda _L^{{ik}} \lambda _L^{{kj}}}{\delta ^2}\Bigg(\Big(-x_Z^3
-x_Z^2 \left(\eta _{{ij}}+6 \xi_{k}\right)
+2 \xi _{{ij}}^2 x_Z
\Big)\Delta _1\nonumber\\
&+\frac{1}{2 x_i}\Big(\left(\xi _{{ik}}+1\right) x_Z^3
+x_Z^2 \left(x_i \left(7 \xi_{k}-x_j\right)+3 x_j \xi_{k}+3 x_i^2\right)
+\xi _{{ij}} x_Z \left(x_i \left(x_j+7\xi_{k}\right)+3 x_j \xi_{k}-3 x_i^2\right)
\nonumber\\&
+\xi _{{ij}}^3 \left(1-\eta _{{ik}}\right)
\Big)\Delta _2+\frac{1}{2 x_i x_j}\Big(x_Z^3 \left(2 x_i x_j-\eta _{{ij}} \xi_{k}\right)
+x_Z^2 \left(x_i^2 \left(2 x_j+3\xi_{k}\right)+2 x_i x_j \left(x_j+7\xi_{k}\right)+3 x_j^2 \xi_{k}\right)
\nonumber\\&-\xi _{{ij}}^2 x_Z \left(3 \eta _{{ij}} \xi_{k}+4 x_i x_j\right)
+\xi _{{ij}}^4 \xi_{k}
\Big)\Delta _3+\Big(x_Z^3 \left(2-\eta _{{ij}}-4 x_k\right)
+2 x_Z^2 \left(\eta _{{ij}}\eta_{k}+x_i x_j+\xi_{ij}^2-3 \xi_{k}^2\right)\nonumber\\
&
+\xi _{{ij}}^2 x_Z \left(2x_k-\eta _{{ij}}-4\right)
\Big)C_{ZSkS}-\delta x_Z\Bigg),
\end{align}

\begin{align}
L_3^{'(\rm b)}&=\frac{g_{{VSS}} \sqrt{x_k}}{\delta}\left(\sqrt{x_i}\lambda _L^{{ik}} \lambda _R^{{kj}}+\sqrt{x_j}\lambda _L^{{kj}} \lambda _R^{{ik}}
\right)\Bigg(2 x_Z\Delta_0-\Big(x_Z+\xi _{{ij}}
\Big)\Delta _2
+x_Z\Big( x_Z+ 2 \xi_{k}-\eta _{{ij}}\Big)C_{ZSkS}\Bigg).
\end{align}
Finally  Feynman diagrams (c) and (d) only contribute to monopole terms. The corresponding contribution of both diagrams is
\begin{align}
\sum_{i=1}^{3} L_i^{'(\rm cd)}&=\frac{g_L^j \lambda _R^{{ik}} \lambda _R^{{kj}}}{2\xi _{{ij}}}\Bigg(\Big(\eta _{{ik}}-1
\Big)\Delta _2+\xi _{{ij}}\Delta _3-2x_k\Delta _4
- 2\xi_{k}\Delta _5+ \xi _{{ij}}B_{{kS}}-2\xi_k\Bigg)\nonumber\\&
+\frac{g_L^j \lambda _L^{{ik}} \lambda _L^{{kj}}}{2 \sqrt{x_i} \sqrt{x_j}}\Bigg(\frac{1}{\xi _{{ij}}}\left(\eta _{{ik}}-1\right) x_j
\Delta _2 - \xi_{k}\Delta _3\Bigg)+\frac{
 \sqrt{x_k} g_L^j}{\xi _{{ij}}} \left(\sqrt{x_i} \lambda _L^{{ik}} \lambda _R^{{kj}}+\sqrt{x_j} \lambda _L^{{kj}} \lambda _R^{{ik}}\right)\Delta _2.
\end{align}
We can observe that the ultraviolet divergent term $B_{kS}\equiv B_0(0,m_k^2,m_S^2)$, which appears only in the monopole terms, is canceled  out when summing over all the contributions.

Furthermore, the form factors associated with the right-handed terms are given by
\begin{equation}
R^{Z}=L^Z\left(\lambda^{lm}_L\leftrightarrow \lambda^{lm}_R,g_{ZSS}\to -g_{ZSS}\right),
\end{equation}
and
\begin{equation}
R^{'Z}=-L^{'Z}\left(\lambda^{lm}_L\leftrightarrow \lambda^{lm}_R,g_{ZSS}\to -g_{ZSS}\right).
\end{equation}
\subsection{Feynman parameter results}
\subsubsection{$f_i\to f_j \gamma$ decay}
\label{fitofjgammaCoefficientsFP}
The $L^\gamma$ form factor of Eq. \eqref{MfitofjV} is ultraviolet finite and is given in terms of Feynman parameter integrals as follows
\begin{align}
\label{LgammaFP}
L^\gamma &=\frac{N_cg^2e\sqrt{x_i}}{62c_W^2\pi^2}\frac{1}{2}\int_0^1 dx \int_0^{1-x}dy\Bigg(\frac{Q_k}{\zeta _1} \left(x y \sqrt{x_i} \lambda _L^{{ik}} \lambda _L^{{kj}}-\lambda _R^{{ik}} \left(x \sqrt{x_j} (x+y-1) \lambda _R^{{kj}}+(x-1) \sqrt{x_k} \lambda _L^{{kj}}\right)\right)\nonumber\\&
+\frac{Q_S}{\zeta _2} (x+y-1) \left(x \sqrt{x_i} \lambda _L^{{ik}} \lambda _L^{{kj}}+y \sqrt{x_j} \lambda _R^{{ik}} \lambda _R^{{kj}}+\sqrt{x_k} \lambda _L^{{kj}} \lambda _R^{{ik}}\right)\Bigg),
\end{align} where
\begin{align}
\zeta_1&=x \left(y \xi_{ji}+x_j (x-1)-\xi_k\right)+x_k,\\
\zeta_2&=x y \eta_{ij}-x \left(\eta_{ik}-1\right)+x^2 x_i-y \left(\eta_{jk}-y x_j\right)+x_k+y.
\end{align}

The $R^\gamma$ form factor is given by Eq. \eqref{Rgamma}, whereas  monopole terms $L^{'\gamma}$ and $R{'\gamma}$ are zero as already mentioned (one must consider electric charge conservation).

\subsubsection{$f_i\to f_j Z$ decay}

The dipole terms of Eq. \eqref{LZ}, which only arise from diagrams (a) and (b), are ultraviolet finite and are given by
\begin{equation}
L^{\rm (a)}=\int_0^1 dx\int_0^{1-x}dy\frac {1}{\zeta'_1} \left(x y \sqrt{x_i} g_L^k \lambda _L^{{ik}} \lambda _L^{{kj}}-x \sqrt{x_j} (x+y-1) g_R^k \lambda _R^{{ik}} \lambda _R^{{kj}}+\sqrt{x_k} \lambda _L^{{kj}} \lambda _R^{{ik}} \left(y g_L^k-(x+y-1) g_R^k\right)\right),
\end{equation}

\begin{equation}
L^{\rm (b)}=g_{{VSS}}\int_0^1 dx\int_0^{1-x}dy\frac{(x+y-1) }{\zeta'_2}\left(x \sqrt{x_i} \lambda _L^{{ik}} \lambda _L^{{kj}}+y \sqrt{x_j} \lambda _R^{{ik}} \lambda _R^{{kj}}+\sqrt{x_k} \lambda _L^{{kj}} \lambda _R^{{ik}}\right),
\end{equation}
whereas the partial contributions to the monopole terms of Eq. \eqref{LZ'} are ultraviolet divergent and read
\begin{align}
L^{'\rm (a)}&=\int_0^1 dx\int_0^{1-x}dy\frac{1}{\zeta'_1}\Bigg(x (y-1) \sqrt{x_i} \sqrt{x_j} g_L^k \lambda _L^{{ik}} \lambda _L^{{kj}}+(y-1) \sqrt{x_i} \sqrt{x_k} g_L^k \lambda _L^{{ik}} \lambda _R^{{kj}}\nonumber\\&
+\lambda _R^{{ik}} \lambda _R^{{kj}} \left(g_R^k \left((x+y-1) \left(y (x_Z-\xi _{{ij}})+x x_j\right)+\zeta'_1 \left(1+  \log \left(\zeta'_1\right)\right)  \right)-x_k g_L^k\right)\nonumber\\&+\sqrt{x_j} \sqrt{x_k}\lambda _R^{{ik}} \lambda _L^{{kj}} \left((x+y-1) g_R^k-x g_L^k\right)
\Bigg)- \frac{g_R^k}{2} \lambda _R^{{ik}}\lambda _R^{{kj}}\Delta_{\rm UV},
\end{align}
\begin{align}
L^{'\rm (b)}&=\frac{g_{{VSS}}}{2\zeta'_2}\int_0^1 dx\int_0^{1-x}dy(2 x-1) \Bigg(\sqrt{x_i} \lambda _L^{{ik}} \left(\sqrt{x_j} (x+y) \lambda _L^{{kj}}+\sqrt{x_k} \lambda _R^{{kj}}\right)\nonumber\\&+\lambda _R^{{ik}}
\left(\left(x x_i+y x_j+\frac{2\zeta'_2}{2x-1}  \log \left( \zeta'_2\right)\right) \lambda _R^{{kj}}+\sqrt{x_j} \sqrt{x_k} \lambda _L^{{kj}}\right) \Bigg)-\frac{g_{{VSS}}}{2}\lambda _R^{{ik}}\lambda _R^{{kj}}\Delta_{\rm UV},
\end{align}
\begin{align}
L^{'\rm (cd)}&=\int_0^1 dx\frac{g_L^j}{\xi _{{ij}}} \Bigg( \left(\sqrt{x_i} \lambda _L^{{ik}} \left(x \sqrt{x_j} \lambda _L^{{kj}}+\sqrt{x_k} \lambda _R^{{kj}}\right)+\sqrt{x_j} \sqrt{x_k} \lambda _L^{{kj}} \lambda _R^{{ik}}\right)\left(\log \left( \zeta'_{32}\right)-\log \left( \zeta'_{31}\right)\right)\\
\nonumber&+x \lambda _R^{{ik}} \lambda _R^{{kj}}\left(x_j\log \left( \zeta'_{32}\right)-x_i\log \left( \zeta'_{31}\right)\right)
\Bigg)+ \frac{g_L^j}{2} \lambda _R^{{ik}}\lambda _R^{{kj}}\Delta_{\rm UV}.
\end{align}
where  $\Delta_{\rm UV}$ stands for the ultraviolet divergence, which cancels out when summing over the partial contributions as it is proportional to $g_L^k-g_R^j-g_{VSS}$. We also have defined the following functions

\begin{align}
\zeta'_1&=x y \left(x_Z-\xi_{{ij}}\right)+x^2 x_j-x\left(\eta_{{jk}}-1\right)+x_k+(y-1)y x_Z,\nonumber\\
\zeta'_2&=x^2 x_i+x y \left(\eta_{{ij}}-x_Z\right)-x\left(\eta_{{ik}}-1\right)-y\left(\eta _{{jk}}-yx_j\right)+x_k+y,\nonumber\\
\zeta'_{3a}&=x^2 x_a-x \left(\eta_{{ak}}-1\right)+x_k.
\end{align}
\section{Loop integrals for  the  $f_{i}\to f_j H$ decay}
\label{fitofjHCoefficients}
 The $F_L$ and $F_R$ form factors of Eq. \eqref{fitofjHdecaywidth} are given by

\begin{equation}
\label{fL}
F_{L,R}= \frac{N_cg m_S}{32\pi^2 m_W} \sum_{k={\rm (a), (b), (c)}}^3f_{L,R}^{(k)}+\frac{3}{16\pi^2}\frac{\lambda_S v}{m_S}f_{L,R}^{\rm (d)},
\end{equation}
with $f_{L,R}^{(k)}$ ($\rm k=a,b,c,d$)  being the contributions of the Feynman diagram analogue to the diagram (k) of Fig. \ref{fitofjVdecay}, with the $V$ gauge boson replaced by the Higgs boson. Again we present our results in terms of Passarino-Veltman scalar functions and Feynman parameter integrals.

\subsection{Passarino-Veltman results}
The sum of the contributions of the triangle and bubble diagrams (a), (b) and (c) is ultraviolet finite and reads
\begin{align}
\sum_{k={\rm (a), (b), (c)}}f_L^{(k)}&=\frac{1}{2\chi }\Bigg(\sqrt{x_i}\Big(2    x_k (x_H+ \xi_{ji}) (\xi_{Hj}-x_i-2\xi_{k})C'_1+ (\zeta x_{j} x_i(x_{j} +\xi_k)-8  x_{j}  x_k)\Delta'_1\nonumber\\&
+ (4 x_k (x_i-\xi_{Hj})-\zeta x_{j}^2  (x_i+\xi_k))\Delta'_2+  \zeta x_{j} \xi_{ji}\xi_k\Delta'_3 \Big)\lambda_L^{kj} \lambda_L^{ik}\nonumber\\&
+\sqrt{x_{j}}\Big(2   x_k (x_H-\xi_{ji}) (\xi_{Hj}-x_i-2 \xi_k)C'_1+ (4  x_k (x_i-\xi_{Hj})+\zeta  x_i^2 (x_{j} +\xi_k))\Delta'_1\nonumber\\&
- (\zeta x_{j} x_i (x_i+\xi_k)+8 x_i x_k)\Delta'_2+ \zeta x_i\xi_{ji}   \xi_k\Delta'_3 \Big)\lambda_R^{kj} \lambda_R^{ik}\nonumber\\&+
2\sqrt{x_k}\Big(  \left(x_k \left(2 \xi_{ji}^2+x_H^2-3 x_H \xi_{ij}\right)+x_H (x_H+x_{j}  (2 x_i-1)-x_i)\right)C'_1\nonumber\\&+(\zeta  x_i- x_{j}  (x_H-\xi_{ji}))\Delta'_1 -(\zeta  x_{j}+  x_i (x_H+\xi_{ji}) )\Delta'_2 \Big)\lambda_L^{kj} \lambda_R^{ik}\nonumber\\&+
2\sqrt{x_{j}} \sqrt{x_i} \sqrt{x_k}\Big( x_H (\xi_{Hj}-x_i-2    \xi_k)C'_1 -  (\xi_{ji}+x_H-\zeta )\Delta'_1  + \left(\xi_{ji}-x_H-\zeta \right)\Delta'_2 \Big)\lambda_R^{kj} \lambda_L^{ik}\Bigg),
\end{align}
whereas  the contribution of triangle diagram (d), which is ultraviolet finite by itself, can be written as
\begin{align}
f_L^{\rm (d)}&=\frac{1}{\chi}\Bigg( \sqrt{x_i}\Big( (x_H (x_{j} +\xi_k)-\xi_{ji} (\xi_{ik}+1))C'_2 -2  x_{j} \Delta'_1 + (x_i-\xi_{Hj})\Delta'_2+(x_H+\xi_{ji})\Delta'_4 \Big)\lambda_L^{kj} \lambda_L^{ik}\nonumber\\&+\sqrt{x_{j} }\Big( (x_H (x_i+\xi_k)+\xi_{ji} (\xi_{ik}+1))C'_2+  (x_i-\xi_{Hj})\Delta'_1-2 x_i\Delta'_2+ (x_H-\xi_{ji})\Delta'_4 \Big)\lambda_R^{kj} \lambda_R^{ik}\nonumber\\&- \sqrt{x_k} \chi C'_2 \lambda_L^{kj} \lambda_R^{ik}\Bigg),
\end{align}
where we have introduced the auxiliary variable $\chi=\xi_{ji}^2+x_H^2-2 x_H \xi_{ij}$,
and $\zeta=\chi /(x_{j} x_i \xi_{ji})$. As for the $C'_{j}$ and $\Delta'_{j}$ functions, they are given by
\begin{align}\label{FunEsc}
\Delta'_1&=B_0(m_{j}^2,m_k^2,m_S^2)-B_0(m_H^2,m_k^2,m_k^2),\\
\Delta'_2&= B_0(m_i^2,m_k^2,m_S^2)- B_0(m_H^2,m_k^2,m_k^2),\\
\Delta'_3&=B_0(0,m_k^2,m_S^2)-B_0(m_H^2,m_k^2,m_k^2),\\
\Delta'_4&= B_0(m_H^2,m_S^2,m_S^2)-B_0(m_H^2,m_k^2,m_k^2),\\
C'_1&=m_S^2 C'_0(m_H^2, m_{j}^2, m_i^2, m_k^2, m_k^2, m_S^2),\\
C'_2&=m_S^2 C'_0(m_H^2, m_{j}^2, m_i^2, m_k^2, m_k^2, m_S^2).
\end{align}
It is thus evident that ultraviolet divergences cancel out.
As far as the right-handed terms are concerned, they obey
\begin{align}
f_R^{\rm (k)}&=f_L^{\rm (k)}(L\leftrightarrow R)\quad(k={\rm a,\, b,\, c,\, d}).
\end{align}

\subsection{Feynman parameter results}
Feynman parametrization yield the following results for the $f_{L,R}^{\rm (k)}$  coefficients:
\begin{align}
f_L^{\rm (a)}&= \sqrt{x_k}\int_{0}^{1}dx\int_{0}^{1-x}dy\frac{1}{\varrho_1}\Bigg(\sqrt{x_i} \sqrt{x_k} (2 y-1)\lambda_L^{kj} \lambda_L^{ik}+\sqrt{x_{j} } \sqrt{x_k} (1-2 (x+ y))\lambda_R^{kj} \lambda_R^{ik}\nonumber\\&
      -\Big(2 \varrho_1 \log \left(\varrho_{1}\right)+x y (\xi_{ji}+x_H)+(x-1) x x_{j} +\varrho_1+x_Hy(y-1)+x_k\Big)\lambda_L^{kj} \lambda_R^{ik}-x \sqrt{x_{j} } \sqrt{x_i}\lambda_R^{kj} \lambda_L^{ik}\Bigg)\nonumber\\&
      +\sqrt{x_k} \lambda_L^{kj} \lambda_R^{ik}\Delta_{\rm UV},
\end{align}
\begin{align}
\sum_{k={\rm (b), (c)}}f_L^{\rm (k)}&=\frac{1}{\xi_{ji}}\int_{0}^{1}dx\Bigg(
      \sqrt{x_{j}}\sqrt{x_i}\left( x\left(\sqrt{x_{j}}\lambda_L^{kj} \lambda_L^{ik}+ \sqrt{x_i } \lambda_R^
      {kj} \lambda_R^{ik}\right)+\sqrt{x_k} \lambda_R^{kj} \lambda_L^{ik}\right) (\log (\varrho_{2j})-\log (\varrho_{2i}))\nonumber\\&+
      \sqrt{x_k} (x_{j}  \log (\varrho_{2j})-x_i \log (\varrho_{2i}))\lambda_L^{kj} \lambda_R^{ik}
      +\Bigg)
      -\sqrt{x_k} \lambda_L^{kj} \lambda_R^{ik}\Delta_{\rm UV},
      \end{align}
where it is evident that the ultraviolet divergence $\Delta_{\rm UV}$  cancels out when summing over the partial contributions. We also use the following auxiliary variables
\begin{align*}
\varrho_{1}&=x^2 x_{j} +x (y (x_H+\xi_{ji})+ \xi_{kj}+1)+x_H (y-1) y+x_k,
\end{align*}
and
\begin{align*}
\varrho_{2a}&=x x_a(x-1) -x (x_k-1)+x_k.
\end{align*}
As far as the contribution of Feynman diagram (d) is concerned, it is given by
\begin{align}
      f_L^{\rm (d)}&=\int_{0}^{1}dx\int_{0}^{1-x}dy\frac{1}{ \varrho_{3}}\left(\sqrt{x_k}\lambda_L^{kj} \lambda_R^{ik}-\sqrt{x_i} (x+y-1)\lambda_L^{kj} \lambda_L^{ik}+\sqrt{x_{j} } y\lambda_R^{kj} \lambda_R^{ik}\right),
\end{align}
with
\begin{equation}
\varrho_{4}=x^2 x_i+x (y (x_H-\xi_{ji})+\xi_{ki}-1)+x_H (y-1) y+1.
\end{equation}

\section{Loop integrals for the $H\to \bar{f}_j f_i$ decay}
\label{Htofifjdecay}
As already mentioned, the form factors $F_L$ and $F_R$ for the $f_i\to f_j H$ decay width are also valid for the $H\to f_j f_i$ decay width  given in   \eqref{Htofifjdecaywidth}. It is interesting to obtain the approximate results in the limit of small $x_{j}$ and $x_i$. In the case of the Passarino-Veltman results  $\Delta'_ 1\to \Delta'_3+O(x_{j})$ and $\Delta'_ 2\to \Delta'_3+O(x_i)$ for small $x_i$ and $x_j$, which means that in the limit of vanishing external fermion masses we have
\begin{align}
\sum_{k={\rm (a), (b), (c)}}f_L^{(k)}&\simeq \lambda_L^{kj} \lambda_R^{ik} \sqrt{x_k}\left( (x_k+1)C'_1-\Delta'_3\right),
\end{align}
and
\begin{align}
f_L^{\rm (d)}&=- \lambda_L^{kj} \lambda_R^{ik}\sqrt{x_k} C'_2,
\end{align}
which means that in this scenario the $H\to f_i f_j$ decay width can be written as
\begin{align}
\label{widthap}
\Gamma(H\to f_i f_j)\simeq \frac{m_H}{32 \pi } \left(|\lambda_L^{kj} \lambda_R^{ik}|^2+|\lambda_R^{kj} \lambda_L^{ik}|^2\right)|f(m_k,m_S,m_i,m_j)|^2.
\end{align}
where
\begin{align}
\label{FL}
   f(m_k,m_S,m_i,m_j)& \simeq  \frac{3m_k}{16\pi^2 m_S} \Bigg(\left( (m_k^2+m_S^2)C_0(m_H^2, 0, 0, m_k^2, m_k^2, m_S^2)-\left(B_0(0,m_k^2,m_S^2)-B_0(m_H^2,m_k^2,m_k^2)\right)\right)\nonumber\\
&- m_S^2 C_0(m_H^2, 0, 0, m_k^2, m_k^2, m_S^2)\Bigg),
\end{align}
This result agrees with the one presented in \cite{Cheung:2015yga,Kim:2018oih,Cai:2017wry,Bauer:2015knc}.

As far as the Feynman parameter results, in the vanishing limit of $m_{j}$ and $m_i$ one can obtain
\begin{align}
      f_L^{\rm (a)}&=-\sqrt{x_k}\int_{0}^{1}dx\int_{0}^{1-x}dy\frac{1}{\varrho'_{1}}\left(2 \varrho'_{1} \log (\varrho'_{1})+x_H y (x+y-1)+\varrho'_{1}+x_k\right)\lambda_L^{kj} \lambda_R^{ik}\nonumber\\&
      +\sqrt{x_k} \lambda_L^{kj} \lambda_R^{ik}\Delta_{\rm UV},
\end{align}
\begin{align}
\sum_{k={\rm (b), (c)}} f_L^{(k)}&=\sqrt{x_k} \int_{0}^{1}dx\log (\varrho'_{2})\lambda_L^{kj} \lambda_R^{ik}- \sqrt{x_k} \lambda_L^{kj} \lambda_R^{ik}\Delta_{\rm UV},
\end{align}
and
\begin{align}
f_L^{\rm (d)}&= \sqrt{x_k}\int_{0}^{1}dx\int_{0}^{1-x}dy
\frac{1}{\varrho'_{3}}\lambda_L^{kj} \lambda_R^{ik},
\end{align}
with
\begin{align}
\varrho'_{1}&=x (x_k+1)+x_H y(x+y-1) +x_k,\\
\varrho'_{2}&=x_k-x (x_k-1),\\
\varrho'_{3}&=x  (x_k-1)+x_H y (x+y-1)+1.
\end{align}

\section{ Lepton anomalous magnetic dipole moment}
\label{AnomalousMDM}

The  $F $ and $G $ functions of Eq. \eqref{afi} read
\begin{eqnarray}
F (z_1,z_2)&=&Q_k F _1(z_1,z_2)+Q_{S} F _2(z_1,z_2),\\
G (z_1,z_2)&=&Q_j G _1(z_1,z_2)+Q_{S_k} G _2(z_1,z_2),
\end{eqnarray}
with the $F_a $  and $G_a $  functions given in terms of Feynman parameter integrals  by
\begin{eqnarray}
\label{Fagfeyn}
F _a(z_1,z_2)=2\int_0^1\frac{(1-x) x \xi_a(x)}{(1-x)(z_2- x z_1)+x}dx,\\
\label{Gagfeyn}
G _a(z_1,z_2)=2\int_0^1\frac{(1-x) \xi_a(x)}{(1-x)(z_2- x z_1)+x}dx,
\end{eqnarray}
where $\xi_1(x)=1-x$  and $\xi_2(x)=x$. The integration is straightforward  in the limit of a  light external fermion and  heavy internal fermion and  LQ: $x_i\ll x_k$
\begin{align}
\label{Fgamma1}
F (x_i\simeq 0,x_k)&= \frac{Q_k}{3(1-x_k)^4}\left(2+3x_k-6x_k^2+ x_k^3+6 x_k \log
(x_k)\right)\nonumber\\&+\frac{Q_S}{3(1-x_k)^4}\left(1-6x_k+3x_k^2+ 2x_k^3-6 x_k^2 \log (x_k)\right).
\end{align}
\begin{align}
\label{Ggamma1}
G (x_i\simeq 0,x_k)&=-\frac{Q_k}{(1-x_k)^3}\left(3-4x_k+x_k^2+2 \log (x_k)\right)\nonumber\\
&+\frac{Q_S}{(1-x_k)^3}\left(1-x_k^2+2 x_k
\log (x_k)\right).
\end{align}

For completeness we also present the results  in terms of Passarino-Veltman scalar
functions:
\begin{eqnarray}
F_1 (z_1,z_2)&=&-
\frac{1}{z_1^2 \zeta(z_1,z_2)}\left(2 z_2 \left(2 z_1+\zeta(z_1,z_2) \right)\Delta_6(z_1,z_2)-2
   \left(z_1 \left(1-z_1+z_2\right)+\zeta(z_1,z_2)
\right)\Delta_7(z_1,z_2)\right.\nonumber\\&+&\left.z_1 \left(4 z_2+\zeta(z_1,z_2) -4\right)+2
   \left(z_2-1\right) \zeta(z_1,z_2) +4 z_1^2\right),\\
F_2 (z_1,z_2)&=&-\frac{1}{z_1^2 \zeta(z_1,z_2) }
\left(2 z_2 \left(z_1
   \left(z_2+1\right)-\left(z_2-1\right){}^2\right)\Delta_6(z_1,z_2)+2
   \left(\left(z_2-1\right){}^2+\left(z_1-2\right)
   z_1\right)\Delta_7(z_1,z_2)\right.\nonumber\\&+&\left.z_1^3+\left(z_2-1\right)
\left(z_2+3\right) z_1-2
   \left(z_2-1\right){}^3\right),\\
G_1 (z_1,z_2)&=&-
\frac{1}{z_1 \zeta(z_1,z_2)}\left(2 \left(z_1^2-\left(2 z_2+1\right) z_1+\left(z_2-1\right)
   z_2\right)\Delta_6(z_1,z_2)+2 \left(z_1-z_2+1\right)\Delta_7(z_1,z_2)\right.\nonumber\\&+&\left.2
   \left(z_1-z_2+1\right){}^2\right),\\
G_2 (z_1,z_2)&=&-
\frac{1}{z_1 \zeta(z_1,z_2) }\left(2 \left(z_1-z_2+1\right) z_2\Delta_6(z_1,z_2)+2
\left(z_1+z_2-1\right)\Delta_7(z_1,z_2)+2
   \left(z_1^2-\left(z_2-1\right){}^2\right)\right),
\end{eqnarray}
with
\begin{eqnarray}
\label{PassVelFun2}
\Delta_6(x,y)&=&B_0(0, y\, m_S^2, y \,m_S^2)- B_0(x\, m_S^2, y \,m_S^2,
m_S^2),\\
\Delta_7(x,y)&=&B_0(0, m_S^2, m_S^2)- B_0(x\, m_S^2, y \,m_S^2, m_S^2),
\end{eqnarray}
and $\zeta(x,y)=\left(1+y-x\right)^2-4 y$.

\twocolumn

\end{document}